\documentclass[12pt]{article}
\usepackage{amsmath,amssymb}
\usepackage{bm}
\usepackage{color}
\usepackage{cite}
\usepackage{mathtools}

\usepackage{here}

\usepackage{comment}

\def\slash#1{\not\!\!#1}

\begin{document}

\title{
\begin{flushright}
\ \\*[-80pt]
\begin{minipage}{0.2\linewidth}
\normalsize
EPHOU-21-001\\*[50pt]
\end{minipage}
\end{flushright}
{\Large \bf
Modular flavor symmetries of three-generation modes on magnetized toroidal orbifolds
\\*[20pt]}}

\author{
~Shota Kikuchi,
~Tatsuo Kobayashi,  and 
~Hikaru Uchida
\\*[20pt]
\centerline{
\begin{minipage}{\linewidth}
\begin{center}
{\it \normalsize
Department of Physics, Hokkaido University, Sapporo 060-0810, Japan} \\*[5pt]
\end{center}
\end{minipage}}
\\*[50pt]}

\date{
\centerline{\small \bf Abstract}
\begin{minipage}{0.9\linewidth}
\medskip
\medskip
\small
We study the modular symmetry on magnetized toroidal orbifolds with Scherk-Schwarz phases.
In particular, we investigate finite modular flavor groups for three-generation modes on  magnetized orbifolds.
The three-generation modes can be the three-dimensional irreducible  representations of 
covering groups and central extended groups of $\Gamma_N$ for $N=3,4,5,7,8,16$, 
that is, covering groups of $\Delta(6(N/2)^2)$ for $N=$ even and central extensions of $PSL(2,\mathbb{Z}_{N})$ for $N=$odd 
with Scherk-Schwarz phases.
We also study anomaly behaviors.
\end{minipage}
}

\begin{titlepage}
\maketitle
\thispagestyle{empty}
\end{titlepage}

\newpage


\section{Introduction}
\label{Intro}

The origin of the flavor structure such as quark and lepton masses and their mixing angles 
is one of the most significant mysteries in particle physics.
Non-Abelian discrete flavor symmetries~\cite{
	Altarelli:2010gt,Ishimori:2010au,Ishimori:2012zz,Hernandez:2012ra,
	King:2013eh,King:2014nza} such as $S_N$, $A_N$, $\Delta(3N^2)$, and $\Delta(6M^2)$ for the three generations of quarks and leptons are attractive candidates to realize the flavor structure.
However, in order to obtain the realistic masses and mixing angles of the quarks and leptons, the complicated vacuum alignment of gauge singlet scalars, the  so-called flavons, is required.

The geometries of compact spaces predicted in higher dimensional theories such as superstring theory can be candidates of the origin of the flavor structure. (See Refs.~\cite{Kobayashi:2006wq,Abe:2009vi}.)
For example, a torus and its orbifold have the complex structure modulus $\tau$, which decides the shape of the torus and the orbifold.
There is the modular symmetry $\Gamma \equiv SL(2,\mathbb{Z})$ as well as $\bar{\Gamma} \equiv SL(2,\mathbb{Z})/\mathbb{Z}_2$ as the geometrical symmetry on a torus and some of orbifolds.
Under the modular transformation, chiral zero-modes on the torus and the orbifolds, corresponding to the flavors of quarks and leptons, are transformed.
That is, the modular symmetry can be regarded as the flavor symmetry.
In addition, Yukawa coupligns as well as higher order couplings can be functions of the modulus $\tau$ and then they also transform under the modular transformation since they can be obtained by overlap integrals of the zero-mode profiles on the torus and the orbifolds.
Instead of  flavons, a vacuum expectation value of the modulus $\tau$ breaks the flavor symmetry, and 
characterizes the flavor structure. 
These features are different from ones in the conventional flavor models.
The modular transformation behavior of zero-modes was investigated in magnetized D-brane models~\cite{Kobayashi:2017dyu,Kobayashi:2018rad,Kobayashi:2018bff,Ohki:2020bpo,Kikuchi:2020frp,Kikuchi:2020nxn,Hoshiya:2020hki} and heterotic orbifold models \cite{Lauer:1989ax,Lerche:1989cs,Ferrara:1989qb,Baur:2019kwi,Nilles:2020nnc}.
(See also Refs.~\cite{Kobayashi:2016ovu,Kariyazono:2019ehj,Kobayashi:2020hoc}.)
In particlular, on magnetized $T^2$ with the magnetic flux $M$, there are $M$-number of chiral zero-modes~\cite{Cremades:2004wa} and in recent work~\cite{Kikuchi:2020frp}, it was shown that the zero-modes with $M=$ even and vanishing Scherk-Schwarz (SS) phases behave as modular forms of weight $1/2$ and then they transform as $M$-dimensional representations of the finite modular subgroup $\widetilde{\Gamma}_{2M}$, which is the quadruple covering group of $\Gamma_{2M}$.
There also exists the modular symmetry on the magnetized $T^2/\mathbb{Z}_2^{\rm (t)}$ twisted orbifold.
The number of zero-modes on the magnetized $T^2/\mathbb{Z}_2^{\rm (t)}$ twisted orbifold was investigated in Refs.~\cite{Abe:2008fi,Abe:2013bca,Abe:2008sx,Abe:2015yva}.
Similarly, in Ref.~\cite{Kikuchi:2020nxn}, it was shown that zero-modes on the magnetized $T^2_1 \times T^2_2$ with the magnetic fluxes $M^{(i)}\ (i=1,2)$ on $T^2_i$ and its orbifolds\footnote{Both of modulus on $T^2_i$, $\tau_i$, are identified each other, i.e.~$\tau_1 = \tau_2 \equiv \tau$. Such moduli identification can be realized by certain three-form fluxes~\cite{Kobayashi:2020hoc} or $\mathbb{Z}_2^{\rm (p)}$ permutation.} behave as modular forms of weight $1$ and they transform under the finite modular subgroup $\Gamma'_{2{\rm lcm}(M^{(1)},M^{(2)})}$, which is the double covering group of $\Gamma_{2{\rm lcm}(M^{(1)},M^{(2)})}$.
The number of zero-modes was investigated in Ref.~\cite{Hoshiya:2020hki}.
The modular transformation for Yukawa couplings has also studied in Ref.~\cite{Hoshiya:2020hki}.
Thus, it is important to study the modular flavor symmetries, particularly in magnetized orbifold models. 

Furthermore, the finite modular subgroups $\Gamma_N$ for $N=2,3,4,5$ are isomorphic to $S_3$, $A_4$, $S_4$, $A_5$, respectively~\cite{deAdelhartToorop:2011re}.
Similarly, $\Gamma'_N$ for $N=3,4,5$ are isomorphic to $T'$, $S'_4$, $A'_5$, respectively~\cite{Liu:2019khw}.
These results are well motivated for realistic model buildings.
In particular, in Ref.~\cite{deAdelhartToorop:2011re}, three-dimensional irreducible  representations 
are studied in the finite modular subgroups and it was shown that three-dimensional irreducible representations appear only in 
the finite modular subgroups: $\Gamma_3 \simeq PSL(2,\mathbb{Z}_3) \simeq A_4$, $\Gamma_4 \simeq S_4$, $\Gamma_5 \simeq PSL(2,\mathbb{Z}_5) \simeq A_5$, $\Gamma_7 \simeq PSL(2,\mathbb{Z}_7)$, $\Gamma_8 \supset \Delta(96)$, and $\Gamma_{16} \supset \Delta(384)$.
Note that a triplet representation of $\Gamma_8$ ($\Gamma_{16}$) is not faithful, 
but represents its subgroup $\Delta(96)$ ($\Delta(384)$) \cite{deAdelhartToorop:2011re}.
Recently, the bottom-up approach of model building with the modular flavor symmetries was 
studied extensively for $\Gamma_N$ \cite{Feruglio:2017spp} and for its covering groups~\cite{Liu:2019khw,Novichkov:2020eep}.

In this paper, we study modular flavor groups of the three-generation modes on magnetized orbifolds.
We study non-vanishing SS phases, although previous studies on the modular symmetry did not include SS phase.
We find that the three-generation modes are the three-dimensional representations of corresponding covering groups and central extended groups of the above finite modular subgroups provided in Ref.~\cite{deAdelhartToorop:2011re}.

After this paper, relevant papers appeared \cite{Almumin:2021fbk,Tatsuta:2021deu}.
In Ref.~\cite{Almumin:2021fbk}, it was claimed that violation of the modular symmetry in 
models with odd magnetic fluxes is strange and it is inconsistent.
To preserve the modular symmetry, a certain shift of the coordinate was introduced 
in the models with odd magnetic fluxes in Ref.~\cite{Almumin:2021fbk}.
That is one class of compacfitication.
However, the modular symmetry can break 
when we impose further boundary conditions on wavefunctions 
by geometry and/or gauge background, that is, a generic compactification.
For example, $T^2/\mathbb{Z}_N$ orbifolds with $N=3,4,6$ break the modular symmetry, 
while some residual symmetries remain.
The full modular symmetry remains in wavefunctions on $T^2$ and $T^2/\mathbb{Z}_2$ 
with even magnetic fluxes and vanishing Wilson lines (WLs), which are equivalent to SS phases.
However, non-vanishing SS phases can break the modular symmetry for even magnetic fluxes.
Indeed, the number of zero-modes depends on SS phases \cite{Abe:2013bca,Abe:2015yva}.
On the other hand, the modular symmetry is broken in wavefunctions for odd magnetic fluxes 
and vanishing Wilson lines and SS phases, but the modular symmetry remains 
for odd magnetic fluxes and non-vanishing WLs, which is a discrete shift of the coordinate.
This result is consistent with Ref.~\cite{Almumin:2021fbk}.
At any rate, a general class of compactifications can be decomposed into two classes.
One class of compactifications 
preserves the modular symmetry, while the other class breaks the modular symmetry.
Both are consistent compactifications.
Thus, one can concentrate on the compactification preserving the modular symmetry, 
or one can discuss generic compactification including breaking of the modular symmetry.
In Ref.~\cite{Tatsuta:2021deu}, SS phases were also studied.

This paper is organized as follows.
In section \ref{T2noSSphase}, we review the modular symmetry on magnetized $T^2$ and $T^2/\mathbb{Z}_2^{{\rm(t)}}$ twisted orbifold without the SS phases.
In section \ref{T2SSphase}, we study the modular symmetry on magnetized $T^2$ and $T^2/\mathbb{Z}_2^{{\rm(t)}}$ twisted orbifold with the SS phases. We can consider the modular symmetry of not only wavefunctions with the magnetic flux $M=$even and the vanishing SS phases but also ones with the magnetic flux $M=$odd and the certain SS phases.
In section \ref{T2three}, we show the specific modular flavor groups for three-generation modes on magnetized $T^2/\mathbb{Z}_2^{{\rm(t)}}$ twisted orbifold with the SS phases. We can find the three-generation modes are the three-dimensional representations of the quadruple covering groups and  $Z_8$ central extended groups of the corresponding modular flavor groups provided in Ref.~\cite{deAdelhartToorop:2011re}.
We also extend the analyses to the modular symmetry on magnetized $T^2_1/\mathbb{Z}_2^{({\rm t}_1)} \times T^2/\mathbb{Z}_2^{({\rm t}_2)}$ orbifold and the $\mathbb{Z}_2^{\rm (p)}$ permutation orbifold, i.e.~$(T^2_1 \times T^2_2)/(\mathbb{Z}_2^{\rm (t)}\times \mathbb{Z}_2^{\rm (p)})$ orbifold in sections \ref{T2T2} and \ref{T2T2three}. We can obtain three-dimensional representations of all the double covering groups of $\Gamma_N$ for $N=4,8,16$, i.e. covering groups of 
$\Delta(6N'^2)$ with $N'=N/2$, 
and $Z_4$ central extended groups of $\Gamma_N$ for $N=3,5,7$, i.e. $Z_4$ extensions of 
$PSL(2,\mathbb{Z}_N)$.
In section \ref{conclusion}, we conclude this study.
In Appendix \ref{SSWL}, we review that the SS phases can be replaced by the WLs through gauge transformation and we show that the modular transformation for them are consistent.
In Appendix \ref{SSshift}, we also show that the $\mathbb{Z}_N$ SS phases are related to the $\mathbb{Z}_N$ shift modes.
In Appendix \ref{proof}, we prove that $\widetilde{\Delta}(6M^2)$, which is the quadruple covering group of $\Delta(6M^2)$, can be obtained.
In Appendix \ref{3dmodularform}, we express three-dimensional modular forms obtained from the wavefunctions on magnetized orbifolds.


\section{Modular symmetry on magnetized $T^2$ and $T^2/\mathbb{Z}_2^{{\rm(t)}}$ twisted orbifold without the Scherk-Schwarz phases}
\label{T2noSSphase}

In this section, we review the modular symmetry on magnetized $T^2$ and $T^2/\mathbb{Z}_2^{{\rm(t)}}$ twisted orbifold without the SS phases.

First, we review the moular symmetry of $T^2$~\cite{Gunning:1962,Schoeneberg:1974,Koblitz:1984,Bruinier:2008}.
A two-dimensional torus $T^2$ can be constructed as $T^2 \simeq \mathbb{C}/\Lambda$, where $\Lambda$ is a two-dimensional lattice spanned by lattice vectors $e_k$ ($k=1,2$).
The torus is characterized by the complex structure modulus $\tau \equiv e_2/e_1$ (${\rm Im}\tau>0$).
We also define the complex coordinate of $\mathbb{C}$ as $u$ and one of $T^2$ as $z \equiv u/e_1$, so that $z+1$ and $z+\tau$ are identified with $z$.
The metric on $T^2$ is given by
\begin{align}
  ds^2 = 2h_{\mu\nu}dz^{\mu}d\bar{z}^{\nu}, \quad 
  h = |e_{1}|^2
  \begin{pmatrix}
    0 & \frac{1}{2} \\
    \frac{1}{2} & 0
  \end{pmatrix},
\end{align}
and then the area of $T^2$ is ${\cal A} = |e_{1}|^2 {\rm Im}\tau$.

Here, we can consider the same lattice spanned by the following lattice vectors transformed by $SL(2,\mathbb{Z}) \equiv \Gamma$,
\begin{align}
  \begin{pmatrix}
    e_2' \\
    e_1'
  \end{pmatrix}
  =
  \begin{pmatrix}
    a & b \\
    c & d
  \end{pmatrix}
  \begin{pmatrix}
    e_2 \\
    e_1
  \end{pmatrix}, \quad
  \gamma =
  \begin{pmatrix}
    a & b \\
    c & d
  \end{pmatrix}
  \in SL(2,\mathbb{Z}) \equiv \Gamma.
\end{align}
The $SL(2,\mathbb{Z})$ is generated by
\begin{align}
  S =
  \begin{pmatrix}
    0 & 1 \\
    -1 & 0
  \end{pmatrix}, \quad
  T =
  \begin{pmatrix}
    1 & 1 \\
    0 & 1
  \end{pmatrix},
\end{align}
and they satisfy the following algebraic relations,
\begin{align}
  Z \equiv S^2 = -\mathbb{I}, \quad Z^2 = S^4 = (ST)^3 = \mathbb{I}. \label{STrel}
\end{align}
Under the $SL(2,\mathbb{Z})$ transformation, the complex coordinate of the torus $z$ and the complex structure modulus $\tau$ are transformed as
\begin{align}
  \gamma: \left(z,\tau\right) \rightarrow& \left( \gamma \left(z,\tau \right) \right) = \left(\frac{z}{c\tau+d},\frac{a\tau+b}{c\tau+d}\right) . \label{ztaumod} 
\end{align}
The above transformation for the modulus $\tau$ is called the (imhomogeneous) modular transformation 
and also $\bar{\Gamma}\equiv \Gamma/\{\pm\mathbb{I}\}$ is called the (imhomogeneous) modular group 
since $\tau$ is invariant under $Z=-\mathbb{I}$.

We define the principal congruence subgroup, $\Gamma(N)$, of level $N$ by 
\begin{align}
  \Gamma (N) \equiv \left\{ h=
  \begin{pmatrix}
    a' & b' \\
    c' & d'
  \end{pmatrix}
  \in \Gamma \left|
  \begin{pmatrix}
    a' & b' \\
    c' & d'
  \end{pmatrix}
  \equiv
  \begin{pmatrix}
    1 & 0 \\
    0 & 1
  \end{pmatrix}
  \right.
  ({\rm mod}\ N)
  \right\}.
\end{align}
Then, the modular forms, $f(\tau)$, of the (integral) weight $k$ for $\Gamma(N)$ is 
the holomorphic functions of $\tau$ which transform under the modular transformation in Eq.~(\ref{ztaumod}) as
\begin{align}
  f(\gamma(\tau)) = J_k(\gamma,\tau) \rho(\gamma) f(\tau), \quad 
 J_k(\gamma,\tau) = (c\tau + d)^k, \quad
\gamma(\tau) = \frac{a\tau+b}{c\tau+d}, \quad
\gamma = 
  \begin{pmatrix}
    a & b \\
    c & d
  \end{pmatrix}
  \in \Gamma.
\end{align}
Here, $\rho(\gamma)$ denotes the unitary representation of the quotient group $\Gamma_N' \equiv \Gamma/\Gamma(N)$ satisfying the following algebraic relations,
\begin{align}
 \rho(Z) = \rho(S)^2 = (-1)^k\mathbb{I}, \quad \rho(Z)^2 = &\rho(S)^4 = [\rho(S)\rho(T)]^3 = \mathbb{I}, \quad \rho(Z)\rho(T) = \rho(T)\rho(Z), \label{Gammarep} \\
&\rho(T)^N = \mathbb{I}. \label{Gamma'Nrep}
\end{align}
For even weight $k$, in particular, $\rho(\gamma)$ becomes the unitary representation of the quotient group $\Gamma_N \equiv \bar{\Gamma}/\bar{\Gamma}(N)$, where $\bar{\Gamma}(N) \equiv \Gamma(N)/\{\pm\mathbb{I}\}$ for $N=1,2$~\footnote{Since $Z=-\mathbb{I} \in \Gamma(N)$ for $N=1,2$, $\rho(Z)=\mathbb{I}$ should be satisfied and then the modular weight $k$ should be even.} and $\bar{\Gamma}(N) \equiv \Gamma(N)$ for $N>2$.
Note that $\Gamma_N$ for $N=2$, $3$, $4$, and $5$ are isomorphic to $S_3$, $A_4$, $S_4$, and $A_5$, respectively~\cite{deAdelhartToorop:2011re}, and also $\Gamma'_N$ for $N=3$, $4$, and $5$ are isomorphic to the corresponding double covering groups: $T'$, $S'_4$, and $A'_5$, respectively~\cite{Liu:2019khw}.
In what follows, we review the wavefunctions of $(z,\tau)$ on a magnetized torus and then review their behavior as modular forms under the modular transformation in Eq.~(\ref{ztaumod}).

First, let us review the wavefunctions, particularly the zero-mode wavefunctions of the two-dimensional spinor, on the torus with $U(1)$ magnetic flux~\cite{Cremades:2004wa}.
Here, we do not consider the WLs or the SS phases. In the next section, we will study the case with the non-vanishing SS phases\footnote{The WLs can be replaced by the SS phases~\cite{Abe:2013bca}. We review it and also show the consistency in terms of the modular symmetry in Appendix~\ref{SSWL}}.
The $U(1)$ magnetic flux is given by
\begin{align}
F = \frac{\pi iM}{{\rm Im}\tau} dz \wedge d\bar{z}, 
\end{align}
which satisfies the quantization condition, $(2\pi)^{-1}\int_{T^2}F=M \in \mathbb{Z}$.
This flux is induced by the vector potential,
\begin{align}
A(z) = \frac{\pi M}{{\rm Im}\tau} {\rm Im}\left( \bar{z}dz \right). \label{A}
\end{align}
This vector potential transforms under lattice translations as
\begin{align}
A(z+1) &= A(z) + d\left( \frac{\pi M}{{\rm Im}\tau} {\rm Im}z \right) = A(z) + d\chi_1(z), \label{chi1} \\
A(z+\tau) &= A(z) + d\left( \frac{\pi M}{{\rm Im}\tau} {\rm Im}\bar{\tau}z \right) = A(z) + d\chi_2(z), \label{chi2}
\end{align}
which correspond to $U(1)$ gauge transformation.
Thereby, the two-dimensional spinor with $U(1)$ unite charge $q=1$,
\begin{align}
\psi(z,\tau) =
\begin{pmatrix}
\psi_+(z,\tau) \\ \psi_-(z,\tau)
\end{pmatrix},
\end{align}
should satisfy the following boundary conditions,
\begin{align}
&\psi(z+1,\tau) = e^{i\chi_1(z)} \psi(z,\tau) = e^{\pi iM \frac{{\rm Im}z}{{\rm Im}\tau}} \psi(z,\tau), \label{psiz1} \\
&\psi(z+\tau,\tau) = e^{i\chi_2(z)} \psi(z,\tau) = e^{\pi iM \frac{{\rm Im}\bar{\tau}z}{{\rm Im}\tau}} \psi(z,\tau). \label{psiztau}
\end{align}
Under these boundary conditions, we can solve the zero-mode Dirac equation,
\begin{align}
  i\slash{D}\psi(z,\tau) = 0,
\end{align}
and then only $\psi_+(z,\tau)$ ($\psi_-(z,\tau)$) has $|M|$-number of degenerate zero-modes when $M$ is positive (negative).
In what follows, we consider the positive flux $M$.
The $j$-th zero-mode wavefunction on the torus with the flux $M$ is expressed as
\begin{align}
\psi_{T^2}^{j,M}(z,\tau)
&= \left(\frac{M}{{\cal A}^2}\right)^{1/4} e^{\pi iMz \frac{{\rm Im}z}{{\rm Im}\tau}}
\vartheta
\begin{bmatrix}
\frac{j}{M}\\
0
\end{bmatrix}
(Mz, M\tau), \quad \forall j \in \mathbb{Z}_{M}=\{0,1,2,...,M-1\},
\label{psizero}
\end{align}
where $\vartheta$ denotes the Jacobi theta function defined as
\begin{align}
\vartheta
\begin{bmatrix}
a\\
b
\end{bmatrix}
(\nu, \tau)
=
\sum_{l\in \mathbb{Z}}
e^{\pi i (a+l)^2\tau}
e^{2\pi i (a+l)(\nu+b)}.
\end{align}
We take the following normalization condition,
\begin{align}
  \int_{T^2} dzd\bar{z} \left(\psi^{j,M}_{T^2}(z,\tau)\right)^* \psi^{k,M}_{T^2}(z,\tau) 
  = (2{\rm Im}\tau)^{-1/2} \delta_{j,k}. \label{Normalization}
\end{align}

Now, we can see the wavefunctions for $\forall j$ in Eq.~(\ref{psizero}) behave as modular forms of weight $1/2$~\footnote{See in detail e.g.~\cite{Schoeneberg:1974,shimura,Duncan:2018wbw}.} under the modular transformation in Eq.~(\ref{ztaumod})~\cite{Kikuchi:2020frp} as follows.
In order to see that, we first introduce the double covering group of $\Gamma$,
\begin{align}
  \widetilde{\Gamma} &\equiv \{[\gamma,\epsilon]|\gamma\in\Gamma,\epsilon\in\{\pm 1\}\}.
\end{align}
The generators are given by
\begin{align}
  \widetilde{S} \equiv [S,1], \quad \widetilde{T} \equiv [T,1],
\end{align}
and they satisfy the following algebraic relations,
\begin{align}
  \widetilde{Z} \equiv \widetilde{S}^2, \quad \widetilde{Z}^2=\widetilde{S}^4=(\widetilde{S}\widetilde{T})^3=[\mathbb{I},-1], 
\quad \widetilde{Z}^4=\widetilde{S}^8=(\widetilde{S}\widetilde{T})^6=[\mathbb{I},1]\equiv \mathbb{I}, 
\quad \widetilde{Z}\widetilde{T} = \widetilde{T}\widetilde{Z} .
\end{align}
Note that the modular transformation in Eq.~(\ref{ztaumod}) does not change under replacing $\gamma \in \Gamma$ with $\widetilde{\gamma} \equiv [\gamma,\epsilon] \in \widetilde{\Gamma}$.
We also introduce the congruence subgroup,
\begin{align}
  \widetilde{\Gamma}(N) \equiv \{ [h,\epsilon]\in\widetilde{\Gamma} | h\in\Gamma(N),\epsilon = 1\}.
\end{align}
Then, the modular forms, $f(\tau)$, of the (half integral) weight $k/2$ for $\widetilde{\Gamma}(N)$ transform under the modular transformation as
\begin{align}
  f(\widetilde{\gamma}(\tau))& = \widetilde{J}_{k/2} (\widetilde{\gamma},\tau) \widetilde{\rho}(\widetilde{\gamma}) f(\tau), \quad \widetilde{\gamma} \in \widetilde{\Gamma}, \label{MFhalf} \\
  \widetilde{J}_{k/2}(\widetilde{\gamma},\tau) = &\epsilon^kJ_{k/2}(\gamma,\tau) = \epsilon^k (c\tau+d)^{k/2}, \quad k\in\mathbb{Z},
\end{align}
where $\widetilde{\rho}(\widetilde{\gamma})$ is the unitary representation of the quotient group $\widetilde{\Gamma}_N \equiv \widetilde{\Gamma}/\widetilde{\Gamma}(N)$, which is the double covering group of $\Gamma'_N$, satisfying the following algebraic relations,
\begin{align}
  &\widetilde{\rho}(\widetilde{Z}) = \widetilde{\rho}(\widetilde{S})^2 = e^{\pi ik/2} \mathbb{I}, \label{Z} \\
  &\widetilde{\rho}(\widetilde{Z})^2 = \widetilde{\rho}(\widetilde{S})^4 = [\widetilde{\rho}(\widetilde{S})\widetilde{\rho}(\widetilde{T})]^3 = e^{\pi ik} \mathbb{I}, \label{Z2} \\
  &\widetilde{\rho}(\widetilde{Z})^4 =\widetilde{\rho}(\widetilde{S})^8 = [\widetilde{\rho}(\widetilde{S})\widetilde{\rho}(\widetilde{T})]^6 = \mathbb{I}, \label{Z4} \\
  &\widetilde{\rho}(\widetilde{Z})\widetilde{\rho}(\widetilde{T}) = \widetilde{\rho}(\widetilde{T})\widetilde{\rho}(\widetilde{Z}) \label{ZT} \\
  &\widetilde{\rho}(\widetilde{T})^N = \mathbb{I}. \label{TN}
\end{align}
Here, we take $(-1)^{k/2}=e^{-\pi ik/2}$.
On the other hand, the wavefunctions for $\forall j$ in Eq.~(\ref{psizero}) transform under the modular transformation as
\begin{align}
  \psi^{j,M}_{T^2}(\widetilde{\gamma}(z,\tau)) &= \widetilde{J}_{1/2}(\widetilde{\gamma}, \tau) \sum_{k=0}^{M-1} \widetilde{\rho}_{T^2}(\widetilde{\gamma})_{jk} \psi^{k,M}_{T^2}(z,\tau), \quad \widetilde{\gamma} \in \widetilde{\Gamma}, \label{wavemodularform} \\
  \widetilde{\rho}_{T^2}(\widetilde{S})_{jk} =& e^{\pi i/4} \frac{1}{\sqrt{M}} e^{2\pi i\frac{jk}{M}}, \quad
\widetilde{\rho}_{T^2}(\widetilde{T})_{jk} = e^{\pi i\frac{j^2}{M}} \delta_{j,k}, \label{rhoSandT}
\end{align}
where $\widetilde{\rho}_{T^2}(\widetilde{\gamma})$ satisfies Eqs.~(\ref{Z})-(\ref{TN}) with $k/2=1/2$ and $N=2M$ although $\mathbb{I}_{jk}=\delta_{j,k}$ in Eq.~(\ref{Z}) is modified into $\delta_{M-j,k}$, derived from
\begin{align}
\psi_{T^2}^{j,M}(\widetilde{Z}(z,\tau)) = \psi_{T^2}^{j,M}(-z,\tau) = \psi_{T^2}^{M-j,M}(z,\tau). \label{-zT2}
\end{align}
Note that the above modular transformation for the wavefunctions without the SS phases can be valid only if the magnetic flux $M$ is even because of the consistency of the boundary conditions in Eqs.~(\ref{psiz1}) and (\ref{psiztau}) under the $T$ transformation.
That is,
the wavefunctions after the $T$ transformation satisfy
\begin{align}
\psi(z+\tau+1,\tau+1) = e^{\pi iM \frac{{\rm Im}(\bar{\tau}+1)z}{{\rm Im}\tau}} \psi(z,\tau+1), \label{psiztauaftT}
\end{align}
while the wavefunctions before the $T$ transformation satisfy
\begin{align}
\psi(z+\tau+1,\tau) = e^{-\pi iM}e^{\pi iM \frac{{\rm Im}(\bar{\tau}+1)z}{{\rm Im}\tau}} \psi(z,\tau). \label{psiztaubefT}
\end{align}
In the next section, however, we will show that when we take the SS phases into account, we can also consider the modular transformation for wavefunctions with the flux $M=$odd.
Thus, the wavefunctions on $T^2$ with the magnetic flux $M \in 2\mathbb{Z}$ and vanishing the SS phases behave as the modular forms of weight $1/2$ for $\widetilde{\Gamma}(2M)$.
They seem to be a $M$-dimensional representation.
However, they can be reducible representation.
Their concrete flavor symmetry depends on irreducible representations.
For example, they can not be faithful.
Thus, we will study concrete flavor symmetries of zero-modes in the following sections.

Finally, we also review the zero-mode wavefunctions on the magnetized $T^2/\mathbb{Z}_2^{{\rm(t)}}$ twisted orbifold without the SS phases~\cite{Abe:2008fi} and the modular transformation for them~\cite{Kikuchi:2020frp}. (See also Ref.~\cite{Kobayashi:2017dyu,Kobayashi:2018rad}.)
$T^2/\mathbb{Z}_2^{{\rm(t)}}$ twisted orbifold can be obtained by further identifying $\mathbb{Z}_2^{{\rm(t)}}$ twisted point $-z$ with $z$.
Note that the modulus $\tau$ is not restricted by $\mathbb{Z}_2^{{\rm(t)}}$ twist orbifolding, which means we can also consider the modular transformation on the $T^2/\mathbb{Z}_2^{{\rm(t)}}$ twisted orbifold.
Then, the wavefunctions on the magnetized $T^2/\mathbb{Z}_2^{{\rm(t)}}$ twisted orbifold should also satisfy the following boundary condition,
\begin{align}
\psi^{j,M}_{T^2/\mathbb{Z}_2^{{\rm(t)}m}}(-z,\tau)  = (-1)^m \psi^{j,M}_{T^2/\mathbb{Z}_2^{{\rm(t)}m}}(z,\tau), \quad m \in \mathbb{Z}_2^{{\rm(t)}}, \label{-zT2Z2}
\end{align}
in addition to the boundary conditions on the magnetized $T^2$ in Eqs.~(\ref{psiz1}) and (\ref{psiztau}).
Actually, their boundary conditions are satisfied by the following linear combination of the wavefunctions on the magnetized $T^2$ as
\begin{align}
\psi^{j,M}_{T^2/\mathbb{Z}_2^{{\rm(t)}m}}(z,\tau) =  {\cal N}_{{\rm(t)}}^j \left(\psi_{T^2}^{j,M}(z,\tau) + (-1)^m \psi_{T^2}^{j,M}(-z,\tau)\right),
\end{align}
where $ {\cal N}_{{\rm(t)}}^j$ denotes the normalization factor determined by the normalization condition in Eq.~(\ref{Normalization}).
Since the wavefunctions on the $T^2$ without the SS phases satisfy Eq.~(\ref{-zT2}),
ones on the $T^2/\mathbb{Z}_2^{{\rm(t)}}$ twisted orbifold without the SS phases can be expanded by
\begin{align}
\psi^{j,M}_{T^2/\mathbb{Z}_2^{{\rm(t)}m}}(z,\tau) =  {\cal N}_{{\rm(t)}}^j \sum_{k=0}^{M-1} \left(\delta_{j,k} + (-1)^m \delta_{M-j,k} \right) \psi_{T^2}^{k,M}(z,\tau),
  \quad {\cal N}_{{\rm(t)}}^j = \left\{
  \begin{array}{l}
    1/2 \quad (j=0,M/2) \\
    1/\sqrt{2} \quad ({\rm otherwise})
  \end{array}
  \right.. \label{psitwist}
\end{align}
In this case without the SS phases, there are $M/2+1$-number of $\mathbb{Z}_2^{{\rm(t)}}$-even ($m=0$) modes and $M/2-1$-number of $\mathbb{Z}_2^{{\rm(t)}}$-odd ($m=1$) modes for $M \in 2\mathbb{Z}$.
Furthermore, under the modular transformation, these transform similarly as Eq.~(\ref{wavemodularform}) replacing Eq.~(\ref{rhoSandT}) with
\begin{align}
\widetilde{\rho}_{T^2/\mathbb{Z}_2^{{\rm(t)}0}} (\widetilde{S})_{jk} &= {\cal N}_{{\rm(t)}}^j {\cal N}_{{\rm(t)}}^k  \frac{4e^{\pi i/4}}{\sqrt{M}} \cos \left(\frac{2\pi jk}{M}\right),\quad 
  \widetilde{\rho}_{T^2/\mathbb{Z}_2^{{\rm(t)}0}}(\widetilde{T})_{jk} = e^{\pi i\frac{j^2}{M}} \delta_{j,k}, \label{rhoSandTZ2even} \\
\widetilde{\rho}_{T^2/\mathbb{Z}_2^{{\rm(t)}1}} (\widetilde{S})_{jk} &= {\cal N}_{{\rm(t)}}^j {\cal N}_{{\rm(t)}}^k  \frac{4ie^{\pi i/4}}{\sqrt{M}} \sin \left(\frac{2\pi jk}{M}\right),\quad 
  \widetilde{\rho}_{T^2/\mathbb{Z}_2^{{\rm(t)}1}}(\widetilde{T})_{jk} = e^{\pi i\frac{j^2}{M}} \delta_{j,k},  \label{rhoSandTZ2odd}
\end{align}
where $\widetilde{\rho}_{T^2/\mathbb{Z}_2^{{\rm(t)}m}}(\widetilde{\gamma})$ for each $m \in \mathbb{Z}_2$ satisfies Eqs.~(\ref{Z})-(\ref{TN}) with $k/2=1/2$ and $N=2M$ although $\mathbb{I}_{jk}=\delta_{j,k}$ in Eq.~(\ref{Z}) is modified into $(-1)^m\delta_{j,k}$, derived from Eq.~(\ref{-zT2Z2}).
Thus, both the $\mathbb{Z}_2^{{\rm(t)}}$-even and odd mode wavefunctions on the $T^2/\mathbb{Z}_2^{{\rm(t)}}$ twisted orbifold with the magnetic flux $M \in 2\mathbb{Z}$ and vanishing the SS phases behave as the modular forms of weight $1/2$.
They decompose into  $(M/2+1)$ and $(M/2-1)$-dimensional representations for $\mathbb{Z}_2^{{\rm(t)}}$-even and odd modes, respectively.
That is, the representations on the magnetized $T^2$ can be decomposed into smaller representations on the magnetized $T^2/\mathbb{Z}_2^{{\rm(t)}}$ twisted orbifold.
We will study their concrete flavor symmetries in the following sections.


\section{Modular symmetry on magnetized $T^2$ and $T^2/\mathbb{Z}_2^{{\rm(t)}}$ twisted orbifold with the Scherk-Schwarz phases}
\label{T2SSphase}

In this section, we review the wavefunctions on magnetized $T^2$ and $T^2/\mathbb{Z}_2^{{\rm(t)}}$ twisted orbifold with the SS phases~\cite{Abe:2013bca} and then we study the modular symmetry for them.

The wavefunctions on $T^2$ with the flux $M$ and the SS phases $(\alpha_1,\alpha_2)$ ($0 \leq \alpha_1, \alpha_2 < 1$)~\footnote{The wavefunction on the magnetized $T^2 \simeq \mathbb{C}/\Lambda$ with the $\mathbb{Z}_N$ SS phases is related to the $\mathbb{Z}_N$-eigenmode wavefunction on the magnetized $\mathbb{Z}_N$ full shifted orbifold of $\widetilde{T}^2 \simeq \mathbb{C}/\widetilde{\Lambda} \ (\widetilde{\Lambda}=N\Lambda)$ withouth the SS phases~\cite{Kikuchi:2020frp,Fujimoto:2013xha}, as shown in Appendix~\ref{SSshift}. The analyses for the wavefunctions on the magnetized $T^2$ with the ($\mathbb{Z}_N$) SS phases are consistent with ones for the wavefunctions on the magnetized $\widetilde{T}^2/\mathbb{Z}_N$ full shifted orbifold without the SS phases in Ref.~\cite{Kikuchi:2020frp}.} satisfy the following boundary conditions,
\begin{align}
&\psi^{\alpha_1,\alpha_2}(z+1,\tau) = e^{2\pi i \alpha_1}e^{i\chi_1(z)} \psi^{\alpha_1,\alpha_2}(z,\tau) = e^{2\pi i \alpha_1}e^{\pi iM \frac{{\rm Im}z}{{\rm Im}\tau}} \psi^{\alpha_1,\alpha_2}(z,\tau), \label{psiz1SS} \\
&\psi^{\alpha_1,\alpha_2}(z+\tau,\tau) = e^{2\pi i \alpha_2}e^{i\chi_2(z)} \psi^{\alpha_1,\alpha_2}(z,\tau) =  e^{2\pi i \alpha_2}e^{\pi iM \frac{{\rm Im}\bar{\tau}z}{{\rm Im}\tau}} \psi^{\alpha_1,\alpha_2}(z,\tau), \label{psiztauSS}
\end{align}
instead of Eqs.~(\ref{psiz1}) and(\ref{psiztau}).
Then, the $j$-th zero-mode wavefunction is expressed as
\begin{align}
\psi_{T^2}^{(j+\alpha_1,\alpha_2),M}(z,\tau)
&= \left(\frac{M}{{\cal A}^2}\right)^{1/4} e^{\pi iMz \frac{{\rm Im}z}{{\rm Im}\tau}}
\vartheta
\begin{bmatrix}
\frac{j+\alpha_1}{M}\\
-\alpha_2
\end{bmatrix}
(Mz, M\tau), \quad \forall j \in \mathbb{Z}_{M}.
\label{psizeroSS}
\end{align}
Note that Eq.~(\ref{psizero}) corresponds to Eq.~(\ref{psizeroSS}) with $(\alpha_1,\alpha_2)=(0,0)$.

Let us study the modular transformation for the wavefunction in Eq.~(\ref{psizeroSS}).
First, we check the consistency of the boundary conditions under the modular transformation.
For example, the wavefunctions after the $T$ transformation satisfy
\begin{align}
\psi^{\alpha'_1,\alpha'_2}(z+\tau+1,\tau+1) = e^{2\pi i \alpha'_2} e^{\pi iM \frac{{\rm Im}(\bar{\tau}+1)z}{{\rm Im}\tau}} \psi^{\alpha_1,\alpha_2}(z,\tau+1), \label{psiztauaftTSS}
\end{align}
while the wavefunctions before the $T$ transformation satisfy
\begin{align}
\psi^{\alpha_1,\alpha_2}(z+\tau+1,\tau) = e^{2\pi i (\alpha_1+\alpha_2-M/2)} e^{\pi iM \frac{{\rm Im}(\bar{\tau}+1)z}{{\rm Im}\tau}} \psi^{\alpha_1,\alpha_2}(z,\tau). \label{psiztaubefTSS}
\end{align}
Thus, in order to see the modular symmetry, particularly the $T$ symmetry, of the wavefunctions, $\alpha'_2 \equiv \alpha_1+\alpha_2-M/2 \ ({\rm mod}\ 1)$ should be satisfied. Also, $\alpha'_1 \equiv \alpha_1 \ ({\rm mod}\ 1)$ is required under the $T$ transformation.
Under the $S$ transformation, similarly, $\alpha'_2 \equiv \alpha_1 \ ({\rm mod}\ 1)$, $\alpha'_1 \equiv 1-\alpha_2 \ ({\rm mod}\ 1)$ are required.
Then, the modular transformation in Eqs.~(\ref{wavemodularform}) and (\ref{rhoSandT}) are deformed as
\begin{align}
  \psi^{(j+\alpha'_1,\alpha'_2),M}_{T^2}(\widetilde{\gamma}(z,\tau)) &= \widetilde{J}_{1/2}(\widetilde{\gamma}, \tau) \sum_{k=0}^{M-1} \widetilde{\rho}_{T^2}(\widetilde{\gamma})_{jk} \psi^{(k+\alpha_1,\alpha_2),M}_{T^2}(z,\tau), \quad \widetilde{\gamma} \in \widetilde{\Gamma}, \label{wavemodularformSS} \\
  \widetilde{\rho}_{T^2}(\widetilde{S})_{jk} &= e^{\pi i/4} \frac{1}{\sqrt{M}} e^{2\pi i((j+1)k+(1-\alpha'_1)\alpha_1)/M} \delta_{\alpha'_2,\alpha_1} \delta_{1-\alpha'_1,\alpha_2}, \label{rhoSSS} \\
\widetilde{\rho}_{T^2}(\widetilde{T})_{jk} &= e^{\pi i(j+\alpha'_1)(j-\alpha'_1+x)/M} \delta_{j,k} \delta_{\alpha_1,\alpha'_1} \delta_{\alpha'_2-\alpha'_1+x/2,\alpha_2}, \label{rhoTSS}
\end{align}
where $x \equiv M \ ({\rm mod}\ 2)$ and $\widetilde{\rho}_{T^2}(\widetilde{\gamma})$ satisfies Eqs.~(\ref{Z})-(\ref{ZT}) with $k/2=1/2$ although $\mathbb{I}_{jk}$ in Eq.~(\ref{Z}) is modified into $e^{-2\pi i(j+\alpha'_1)/M} \delta_{M-j-1,k} \delta_{1-\alpha'_1,\alpha_1} \delta_{1-\alpha'_2,\alpha_2}$, derived from
\begin{align}
\psi_{T^2}^{(j+\alpha_1,\alpha_2),M}(\widetilde{Z}(z,\tau)) = \psi_{T^2}^{(j+\alpha_1,\alpha_2),M}(-z,\tau) = e^{-2\pi i(j+\alpha_1)/M} \psi_{T^2}^{(M-(j+\alpha_1),1-\alpha_2),M}(z,\tau). \label{-zT2SS}
\end{align}
However, Eq.~(\ref{TN}) is not obtained in the general SS phases.
Note that under the modular transformation, in general, the wavefunctions with the SS phases $(\alpha_1,\alpha_2)$ transform into ones with the different SS phases $(\alpha'_1,\alpha'_2)$.
Conversely, when $M$ is even, only the wavefunctions with $(\alpha_1,\alpha_2)=(0,0)$ are closed under the modular transformation. This case is reviewed in previous section.
Similarly, when $M$ is odd, only the wavefunctions with $(\alpha_1,\alpha_2)=(1/2,1/2)$ are closed under the modular transformation. In this case, $\widetilde{\rho}_{T^2}(\widetilde{T})$ satisfies
\begin{align}
\widetilde{\rho}_{T^2}(\widetilde{T})^{M} = e^{\pi i/4} \mathbb{I}, \quad \widetilde{\rho}_{T^2}(\widetilde{T})^{8M}=\mathbb{I}. \label{T8}
\end{align}
Thus, the wavefunctions on $T^2$ with the magnetic flux $M \in 2\mathbb{Z}+1$ and the SS phases $(\alpha_1,\alpha_2)=(1/2,1/2)$ behave as the modular forms of weight $1/2$.
They transform as $M$-dimensional representations, but they can be reducible.

Furthermore, we consider the magnetized $T^2/\mathbb{Z}_2^{{\rm(t)}}$ twisted orbifold with the SS phases\footnote{Similarly, the wavefunctions on the magnetized $T^2/\mathbb{Z}_2^{{\rm(t)}}$ twisted orbifold with the SS phases are related to ones on the magnetized $\widetilde{T}^2/\mathbb{Z}_2$ twisted and full shifted orbifold without the SS phases in Ref.~\cite{Kikuchi:2020frp}.}.
In this case, we can only consider the $\mathbb{Z}_2$ SS phases, $(\alpha_1,\alpha_2)=(\ell_1/2,\ell_2/2)$, ($\ell_1, \ell_2 \in \mathbb{Z}_2$), which are derived from
\begin{align}
1-\alpha_1 \equiv \alpha_1\ ({\rm mod}\ 1), \quad 1-\alpha_2 \equiv \alpha_2\ ({\rm mod}\ 1). \label{condition}
\end{align}
The wavefunctions on the magnetized $T^2/\mathbb{Z}_2^{{\rm(t)}}$ twisted orbifold with the $\mathbb{Z}_2$ SS phases can be expanded by ones on the magnetized $T^2$ in Eq.~(\ref{psizeroSS}) as
\begin{align}
\psi^{\left(j+\frac{\ell_1}{2},\frac{\ell_2}{2} \right),M}_{T^2/\mathbb{Z}_2^{{\rm(t)}m}}(z,\tau) =  {\cal N}_{{\rm(t)}}^{\left(j+\frac{\ell_1}{2},\frac{\ell_2}{2} \right)} \sum_{k=0}^{M-1} \left(\delta_{j,k} + (-1)^m e^{-2\pi i\left(j+\frac{\ell_1}{2} \right)\ell_2/m} \delta_{M-j-\ell_1,k} \right) \psi_{T^2}^{\left(k+\frac{\ell_1}{2},\frac{\ell_2}{2} \right),M}(z,\tau),
\label{psitwistSS}
\end{align}
where we use Eq.~(\ref{-zT2SS}) instead of Eq.~(\ref{-zT2}).
Then, the modular transformation for the wavefunctions in Eq.~(\ref{psitwistSS}) is similarly obtained by replacing Eqs.~(\ref{rhoSSS}) and (\ref{rhoTSS}) with
\begin{align}
\widetilde{\rho}_{T^2/\mathbb{Z}_2^{{\rm(t)}0}} (\widetilde{S})_{jk} &= {\cal N}_{{\rm(t)}}^{\left(j+\frac{\ell_1}{2},\frac{\ell_2}{2} \right)} {\cal N}_{{\rm(t)}}^{\left(k+\frac{\ell_1}{2},\frac{\ell_2}{2} \right)} \frac{4e^{\pi i/4}}{\sqrt{M}} e^{\pi i(k\ell'_1-j\ell_1)} \cos \left(2\pi \left(j+\frac{\ell'_1}{2}\right)\left(k+\frac{\ell_1}{2}\right)/M\right) \delta_{\ell'_2,\ell_1} \delta_{\ell'_1,\ell_2},\label{rhoSZ2evenSS} \\
  \widetilde{\rho}_{T^2/\mathbb{Z}_2^{{\rm(t)}0}}(\widetilde{T})_{jk} &= e^{\pi i\left(j+\frac{\ell'_1}{2}\right)\left(j-\frac{\ell'_1}{2}+x\right)/M} \delta_{j,k} \delta_{\ell'_1,\ell_1} \delta_{\ell'_2-\ell'_1+x,\ell_2}, \label{rhoTZ2evenSS} \\
\widetilde{\rho}_{T^2/\mathbb{Z}_2^{{\rm(t)}1}} (\widetilde{S})_{jk} &= {\cal N}_{{\rm(t)}}^{\left(j+\frac{\ell_1}{2},\frac{\ell_2}{2} \right)} {\cal N}_{{\rm(t)}}^{\left(k+\frac{\ell_1}{2},\frac{\ell_2}{2} \right)} \frac{4ie^{\pi i/4}}{\sqrt{M}} e^{\pi i(k\ell'_1-j\ell_1)} \sin \left(2\pi \left(j+\frac{\ell'_1}{2}\right)\left(k+\frac{\ell_1}{2}\right)/M\right) \delta_{\ell'_2,\ell_1} \delta_{\ell'_1,\ell_2},\label{rhoTZ2oddSS} \\
  \widetilde{\rho}_{T^2/\mathbb{Z}_2^{{\rm(t)}1}}(\widetilde{T})_{jk} &= e^{\pi i\left(j+\frac{\ell'_1}{2}\right)\left(j-\frac{\ell'_1}{2}+x\right)/M} \delta_{j,k} \delta_{\ell'_1,\ell_1} \delta_{\ell'_2-\ell'_1+x,\ell_2}. \label{rhoTZ2oddSS}
\end{align}
In particular, when $M=$even and $(\alpha_1,\alpha_2)=(0,0)$, they correspond to Eqs.~(\ref{rhoSandTZ2even}) and (\ref{rhoSandTZ2odd}).
When $M=$odd and $(\alpha_1,\alpha_2)=(1/2,1/2)$, they become
\begin{align}
\widetilde{\rho}_{T^2/\mathbb{Z}_2^{{\rm(t)}0}} (\widetilde{S})_{jk} &= {\cal N}_{{\rm(t)}}^{\left(j+\frac{1}{2},\frac{1}{2} \right)} {\cal N}_{{\rm(t)}}^{\left(k+\frac{1}{2},\frac{1}{2} \right)} \frac{4e^{\pi i/4}}{\sqrt{M}} e^{\pi i(k-j)} \cos \left(2\pi \left(j+\frac{1}{2}\right)\left(k+\frac{1}{2}\right)/M\right), \label{rhoSZ2evenSS12} \\
  \widetilde{\rho}_{T^2/\mathbb{Z}_2^{{\rm(t)}0}}(\widetilde{T})_{jk} &= e^{\pi i\left(j+\frac{1}{2}\right)^2/M} \delta_{j,k}, \label{rhoTZ2evenSS12} \\
\widetilde{\rho}_{T^2/\mathbb{Z}_2^{{\rm(t)}1}} (\widetilde{S})_{jk} &= {\cal N}_{{\rm(t)}}^{\left(j+\frac{1}{2},\frac{1}{2} \right)} {\cal N}_{{\rm(t)}}^{\left(k+\frac{1}{2},\frac{1}{2} \right)} \frac{4ie^{\pi i/4}}{\sqrt{M}} e^{\pi i(k-j)} \sin \left(2\pi \left(j+\frac{1}{2}\right)\left(k+\frac{1}{2}\right)/M\right),\label{rhoTZ2oddSS12} \\
  \widetilde{\rho}_{T^2/\mathbb{Z}_2^{{\rm(t)}1}}(\widetilde{T})_{jk} &= e^{\pi i\left(j+\frac{1}{2}\right)^2/M} \delta_{j,k}, \label{rhoTZ2oddSS12}
\end{align}
where Eqs.~(\ref{rhoSZ2evenSS12})-(\ref{rhoTZ2oddSS12}) for each $m \in \mathbb{Z}_2$ satisfy Eqs.~(\ref{Z})-(\ref{ZT}), and (\ref{T8}) with $k/2=1/2$ although $\mathbb{I}_{jk}=\delta_{j,k}$ in Eq.~(\ref{Z}) is modified into $(-1)^m\delta_{j,k}$, derived from Eq.~(\ref{-zT2Z2}).
Note that there are $(M-1)/2$-number of $\mathbb{Z}_2^{{\rm(t)}}$-even ($m=0$) modes and $(M+1)/2$-number of $\mathbb{Z}_2^{{\rm(t)}}$-odd ($m=1$) modes when $M=$odd and $(\alpha_1,\alpha_2)=(1/2,1/2)$.
Thus, both $\mathbb{Z}_2^{{\rm(t)}}$-even and odd mode wavefunctions on the $T^2/\mathbb{Z}_2^{{\rm(t)}}$ twisted orbifold with the magnetic flux $M \in 2\mathbb{Z}+1$ and the SS phases $(\alpha_1,\alpha_2)=(1/2,1/2)$ behave as the modular forms of weight $1/2$.
Then, they transform as $(M-1)/2$ and $(M+1)/2$-dimensional representations for $\mathbb{Z}_2^{{\rm(t)}}$-even and odd modes, respectively.
We show the number of the $\mathbb{Z}_2^{{\rm(t)}}$-eigenmodes, $N_m(M)$, which have the modular symmetry, and the order of 
$\tilde T$, i.e.,$\tilde T^h=\mathbb{I}$ in Tables \ref{even} and \ref{odd}.

\begin{table}[H]
\centering
\begin{tabular}{|c|c|c|c|c|c|} \hline
 & $M$ & 2 & 4 & 6 & 8  \\ \hline
$\mathbb{Z}_2^{{\rm(t)}}$-even: $N_0(M)$ & $\frac{M}{2}+1$ & 2 & \fbox{3} & 4 & 5 \\ \hline
$\mathbb{Z}_2^{{\rm(t)}}$-odd: $N_1(M)$ & $\frac{M}{2}-1$ & 0 & 1 & 2 & \fbox{3} \\ \hline
order $h$ of $\tilde T$ ($T^h=\mathbb{I}$) & ${2M}$ & ${4}$ & ${8}$ & ${12}$ & ${16}$ \\ \hline
\end{tabular}
\caption{The number of the $\mathbb{Z}_2^{{\rm(t)}}$-even ($m=0$) modes, $N_0(M)$, and the $\mathbb{Z}_2^{{\rm(t)}}$-odd ($m=1$) modes, $N_1(M)$, on the $T^2/\mathbb{Z}_2^{{\rm(t)}}$ twisted orbifold with $M=$even and $(\alpha_1,\alpha_2)=(0,0)$, and the order of 
$\tilde T$. The three generations are boxed.}
\label{even}
\centering
\begin{tabular}{|c|c|c|c|c|c|} \hline
 & $M$ & 1 & 3 & 5 & 7  \\ \hline
$\mathbb{Z}_2^{{\rm(t)}}$-even: $N_0(M)$ & $\frac{M-1}{2}$ & 0 & 1 & 2 & \fbox{3} \\ \hline
$\mathbb{Z}_2^{{\rm(t)}}$-odd: $N_1(M)$ & $\frac{M+1}{2}$ & 1 & 2 & \fbox{3} & 4 \\ \hline
order $h$ of $\tilde T$ ($T^h=\mathbb{I}$) & ${8M}$ & ${8}$ & ${24}$ & ${40}$ & ${56}$ \\ \hline
\end{tabular}
\caption{The number of the $\mathbb{Z}_2^{{\rm(t)}}$-even ($m=0$) modes, $N_0(M)$, and the $\mathbb{Z}_2^{{\rm(t)}}$-odd ($m=1$) modes, $N_1(M)$, on the $T^2/\mathbb{Z}_2^{{\rm(t)}}$ twisted orbifold with $M=$odd and $(\alpha_1,\alpha_2)=(1/2,1/2)$, and the order of 
$\tilde T$. The three generations are boxed.}
\label{odd}
\end{table}


\section{Modular flavor groups of three-generation modes on magnetized $T^2/\mathbb{Z}_2^{{\rm(t)}}$ twisted orbifold}
\label{T2three}

As mentioned in introduction, in Ref.~\cite{deAdelhartToorop:2011re}, three-dimensional representations can be obtained from the specific finite modular subgroups: $\Gamma_3 \simeq A_4$, $\Gamma_4 \simeq S_4$, $\Gamma_5 \simeq A_5$, $\Gamma_7 \simeq PSL(2,\mathbb{Z}_7)$, $\Gamma_8 \supset \Delta(96)$, and $\Gamma_{16} \supset \Delta(384)$.\footnote{See Refs.~\cite{Ishimori:2010au,Ishimori:2012zz,Escobar:2008vc} to see the algebraic relations for the generators of each non-Abelian discrete flavor group.}
In this section, we show that the three-generation modes on the magnetized $T^2/\mathbb{Z}_2^{{\rm(t)}}$ twisted orbifold shown in Tables \ref{even} and \ref{odd} are the representations of the corresponding covering or central extended groups of the modular flavor groups.

\subsection{$T^2/\mathbb{Z}_2^{{\rm(t)}}$ twisted orbifold with magnetic flux $M=$even and vanishing Scherk-Schwarz phases}
\label{T2even}

In this subsection, we show the modular flavor groups of the three-generation modes on the $T^2/\mathbb{Z}_2^{{\rm(t)}}$ twisted orbifold with $M=$even and $(\alpha_1,\alpha_2)=(0,0)$.
As shown in Table \ref{even}, the three-generation modes are obtained from the $\mathbb{Z}_2^{{\rm(t)}}$-even modes with $M=4$ and the $\mathbb{Z}_2^{{\rm(t)}}$-odd modes with $M=8$.
In the following, we show they are the representations of $\widetilde{\Delta}(96)$ and $\widetilde{\Delta}(384)$, 
which are subgroups of $\widetilde{\Gamma}_8$ and $\widetilde{\Gamma}_{16}$, respectively, and 
are the quadruple covering groups of $\Delta(96)$ and $\Delta(384)$, respectively.

First, $\Gamma_N$  satisfy
\begin{align}
S^2=(ST)^3=T^N=\mathbf{1}. \label{GammaN}
\end{align}
On the other hand, $\Delta(96) \simeq (Z_4 \times Z'_4) \rtimes Z_3 \rtimes Z_2 \simeq \Delta(48) \rtimes Z_2$ and $\Delta(384) \simeq (Z_8 \times Z'_8) \rtimes Z_3 \rtimes Z_2 \simeq \Delta(192) \rtimes Z_2$ satisfy
\begin{align}
&a^M=a'^M=b^3=c^2=\mathbf{1}, \ (M=4,8), \label{algDelta} \\
&aa'=a'a, \ cbc^{-1}=b^{-1}, \ bab^{-1}=a^{-1}a'^{-1}, \ ba'b^{-1}=a, \ cac^{-1}=a'^{-1}, ca'c^{-1}=a^{-1}, \notag
\end{align}
where $a^{(')}$, $b$, $c$ denote the generators of $Z^{(')}_M \ (M=4,8)$, $Z_3$, $Z_2$, respectively~\cite{Ishimori:2010au,Ishimori:2012zz,Escobar:2008vc}.
In order to obtain $\Delta(96)$ and $\Delta(384)$ from the above algebra (\ref{GammaN}) 
for $N=8$ and 16, respectively, the following relation,
\begin{align}
(S^{-1}T^{-1}ST)^3=\mathbf{1} \label{X}
\end{align}
should be also satisfied.
Actually, we can show that if $S$ and $T$ satisfy Eq.~(\ref{X}) in addition to Eq.~(\ref{GammaN}) for $N=2M, \ M \in 4\mathbb{Z}$, the following generators\footnote{For $N=2M$, $M=2(2s-1)$ $s \in \mathbb{Z}$, similarly, the generators,
$a=ST^2ST^4$, $a'=ST^2S^{-1}T^{-2}$, $b=T^{\frac{M}{2}}ST^{M}$, and $c=ST^{M}ST^{\frac{3}{2}M}$,
satisfy Eq.~(\ref{algDelta}).},
\begin{align}
a=ST^2ST^4, \quad a'=ST^2S^{-1}T^{-2}, \quad b=T^{\frac{M}{2}+3}ST^{M}, \quad c=ST^{M-2}ST^{\frac{3}{2}M-1}, \label{DeltaST}
\end{align}
satisfy Eq.~(\ref{algDelta}) in Appendix \ref{proof}. (See also Ref.~\cite{Toorop:2011jn,deAdelhartToorop:2011re}.)
In other words, we can obtain $\Delta(6M^2)$
from $\Gamma_{2M}$ by satisfying the additional relation in Eq.~(\ref{X}). 
Similarly, $\widetilde{\Gamma}_{2M}$ satisfy Eqs.~(\ref{Z})-(\ref{TN}) with $k/2=1/2$ and $N=2M$.
If Eq.~(\ref{X}) is also satisfied, especially for $M \in 4\mathbb{Z}$, the following generators,
\begin{align}
a=ST^2S^5T^4, \quad a'=ST^2S^{-1}T^{-2}, \quad b=T^{\frac{M}{2}+3}S^{\frac{3}{2}M-1}T^{M}, \quad c=ST^{M-2}ST^{\frac{3}{2}M-1}, \label{DeltatildeST}
\end{align}
satisfy
\begin{align}
&a^M=a'^M=b^3=c^8=\mathbf{1}, \label{algDeltatilde} \\
&aa'=a'a, \ cbc^{-1}=b^{-1}, \ bab^{-1}=a^{-1}a'^{-1}, \ ba'b^{-1}=a, \ cac^{-1}=a'^{-1}, ca'c=a^{-1}, \notag
\end{align}
which means the generators in Eq.~(\ref{DeltatildeST}) are ones of $\widetilde{\Delta}(6M^2) \simeq (Z_M \times Z_M) \rtimes Z_3 \rtimes Z_8 \simeq \Delta(3M^2) \rtimes Z_8$, where $a^{(')}$, $b$, $c$ denote ones of $Z^{(')}_M$, $Z_3$, $Z_8$, respectively. (We give the proof in Appendix \ref{proof}.)
In other words, we can obtain $\widetilde{\Delta}(6M^2)$, especially for $M \in 4\mathbb{Z}$, from $\widetilde{\Gamma}_{2M}$ by satisfying the additional relation in Eq.~(\ref{X}). 

Let us study the case of the three-generation modes on the $T^2/\mathbb{Z}_2^{{\rm(t)}}$ twisted orbifold with $M=4,8$ and $(\alpha_1,\alpha_2)=(0,0)$.
The $S$ and $T$ transformation matrices for the $\mathbb{Z}_2^{{\rm(t)}}$-even modes with $M=4$ are given by
\begin{align}
S = \frac{e^{\pi i/4}}{2}
\begin{pmatrix}
1 & \sqrt{2} & 1 \\
\sqrt{2} & 0 & -\sqrt{2} \\
1 & -\sqrt{2} & 1
\end{pmatrix},
\quad&
T = 
\begin{pmatrix}
1 & \ & \ \\
\ & e^{\pi i/4} & \ \\
\ & \ & -1
\end{pmatrix},
\label{M4ST}
\end{align}
and ones for the $\mathbb{Z}_2^{{\rm(t)}}$-odd modes with $M=8$ are given by
\begin{align}
S = \frac{e^{3\pi i/4}}{2}
\begin{pmatrix}
1 & \sqrt{2} & 1 \\
\sqrt{2} & 0 & -\sqrt{2} \\
1 & -\sqrt{2} & 1
\end{pmatrix},
\quad&
T = e^{\pi i/8}
\begin{pmatrix}
1 & \ & \ \\
\ & e^{3\pi i/8} & \ \\
\ & \ & -1
\end{pmatrix}.
\label{M8ST}
\end{align}
Note that here and hereafter (as well as in section \ref{T2T2three}), we omit $\rho$.
Both of the above $S$ and $T$ matrices are the same forms as
\begin{align}
S = \frac{e^{i\theta_1}}{2}
\begin{pmatrix}
1 & \sqrt{2} & 1 \\
\sqrt{2} & 0 & -\sqrt{2} \\
1 & -\sqrt{2} & 1
\end{pmatrix},
\quad&
T = e^{i\theta_2}
\begin{pmatrix}
1 & \ & \ \\
\ & e^{i\theta_3} & \ \\
\ & \ & -1
\end{pmatrix}, \quad \forall \theta_{1,2,3} \in \mathbb{R},
\label{STingen}
\end{align}
and we can check that Eq.~(\ref{STingen}) satisfies Eq.~(\ref{X}) in general.
Thus, the three-generation $\mathbb{Z}_2^{{\rm(t)}}$-even modes with $M=4$ and $\mathbb{Z}_2^{{\rm(t)}}$-odd modes with $M=8$ are transformed under the modular transformation as the three-dimensional representations of $\widetilde{\Delta}(96)$ and $\widetilde{\Delta}(384)$, respectively\footnote{See also Ref. \cite{Kobayashi:2018bff}.}.

We also comment on the modular flavor anomaly.
As discussed in Ref.~\cite{Araki:2008ek,Kariyazono:2019ehj}, the transformation $g$ can be anomalous if ${\rm det}(g) \neq 1$.
Then, let us see the anomaly of the modular flavor group $\widetilde{\Delta}(6M^2)$.
From Eqs.~(\ref{Z})-(\ref{TN}) with $k/2=1/2$ and $N=2M$, (\ref{DeltatildeST}), and (\ref{algDeltatilde}), we can obtain
\begin{align}
{\rm det}(a) = {\rm det}(a') = {\rm det}(b) = 1, \ {\rm det}(c) = {\rm det}(T)^{\frac{M}{2}+3}, \ {\rm det}(c)^8 =1. \label{detDeltatilde}
\end{align}
Actually, both of Eqs.~(\ref{M4ST}) and (\ref{M8ST}) satisfy Eq.~(\ref{detDeltatilde}) and ${\rm det}(c)=e^{\pi i/4}$.
Thus, only $Z_8$ symmetry, generated by $c$, can be anomalous\footnote{The anomalous symmetry which is the discrete subsymmetry of $U(1)$ can be cancelled by the Green-Schwarz mechanism.} and then $\Delta(48)$ and $\Delta(192)$ remain anomaly-free, respectively.


\subsection{$T^2/\mathbb{Z}_2^{{\rm(t)}}$ twisted orbifold with magnetic flux $M=$odd and the Scherk-Schwarz phases $(\alpha_1,\alpha_2)=(1/2,1/2)$}
\label{T2odd}

In this subsection, we show the modular flavor groups of the three-generation modes on the $T^2/\mathbb{Z}_2^{{\rm(t)}}$ twisted orbifold with $M=$odd and $(\alpha_1,\alpha_2)=(1/2,1/2)$.
As shown in Table \ref{odd}, the three-generation modes are obtained from the $\mathbb{Z}_2^{{\rm(t)}}$-odd modes with $M=5$ and the $\mathbb{Z}_2^{{\rm(t)}}$-even modes with $M=7$.

First, the $S$ and $T$ transformation matrices for the $\mathbb{Z}_2^{{\rm(t)}}$-odd modes with $M=5$ are given by
\begin{align}
S = \frac{ie^{\pi i/4}}{\sqrt{5}}
\begin{pmatrix}
2 \sin \left(\frac{\pi}{10}\right) & 2 e^{\pi i/5} \sin \left(\frac{3\pi}{10}\right) & \sqrt{2} e^{2\pi i/5} \\
2 e^{-\pi i/5} \sin \left(\frac{3\pi}{10}\right) & 2 \sin \left(\frac{\pi}{10}\right) & - \sqrt{2} e^{\pi i/5} \\
\sqrt{2} e^{-2\pi i/5} & - \sqrt{2} e^{-\pi i/5} & 1
\end{pmatrix}, \ 
T = 
\begin{pmatrix}
e^{\pi i/20} & \ & \ \\
\ & e^{9\pi i/20} & \ \\
\ & \ & e^{25\pi i/20}
\end{pmatrix},
\label{M5ST}
\end{align}
which satisfy Eqs.~(\ref{Z})-(\ref{ZT}) and (\ref{T8}) with $k/2=1/2$ and replacing $\mathbb{I}$ in Eq.~(\ref{Z}) with $(-1)^{m=1}\mathbb{I}=-\mathbb{I}$.
When we define the following generators,
\begin{align}
a=ST^5, \ b=ST^{21}, \ c=T^5, \label{M5gen}
\end{align}
from the above $S$ and $T$ in Eq.~(\ref{M5ST}),
they satisfy
\begin{align}
a^2=b^3=(ab)^5=c^8=\mathbf{1}, \ ac=ca, \ bc=cb, \label{M5alg}
\end{align}
which mean they are the generators of $A_5 \times Z_8$.
Thus, the three-generational $\mathbb{Z}_2^{{\rm(t)}}$-odd modes with $M=5$ are transformed under the modular transformation as the three-dimensional representations of $A_5 \times Z_8$.

Next, the $S$ and $T$ transformation matrices for the $\mathbb{Z}_2^{{\rm(t)}}$-even modes with $M=7$ are given by
\begin{align}
S = \frac{2e^{\pi i/4}}{\sqrt{7}}
\begin{pmatrix}
\cos \left(\frac{\pi}{14}\right) & e^{\pi i/7} \cos \left(\frac{3\pi}{14}\right) & e^{2\pi i/7} \cos \left(\frac{5\pi}{14}\right) \\
e^{-\pi i/7} \cos \left(\frac{3\pi}{14}\right) & \cos \left(\frac{9\pi}{14}\right) & - e^{\pi i/7} \cos \left(\frac{\pi}{14}\right) \\
e^{-2\pi i/7} \cos \left(\frac{5\pi}{14}\right) & - e^{-\pi i/7} \cos \left(\frac{\pi}{14}\right) & \cos \left(\frac{3\pi}{14}\right)
\end{pmatrix}, \ 
T = 
\begin{pmatrix}
e^{\pi i/28} & \ & \ \\
\ & e^{9\pi i/28} & \ \\
\ & \ & e^{25\pi i/28}
\end{pmatrix}
\label{M7ST}
\end{align}
which satisfy Eqs.~(\ref{Z})-(\ref{ZT}) and (\ref{T8}) with $k/2=1/2$.
They also satisfy
\begin{align}
(S^{-1}T^{-1}ST)^4=\mathbf{1}. \label{add}
\end{align}
When we define the following generators,
\begin{align}
a=ST^{21}, \ b=S^7T^3, \ c=T^7 \label{M7gen}
\end{align}
from the above $S$ and $T$ in Eq.~(\ref{M7ST}),
they satisfy
\begin{align}
a^2=b^4=(ab)^7=(a^{-1}b^{-1}ab)^4=c^8=\mathbf{1}, \ ac=ca, \ bc=cb \label{M7alg}
\end{align}
which mean they are the generators of $PSL(2,Z_7) \times Z_8$.
Thus, the three-generational $\mathbb{Z}_2^{{\rm(t)}}$-even modes with $M=7$ are transformed under the modular transformation as the three-dimensional representations of $PSL(2,Z_7) \times Z_8$.

Similarly, we comment on the anomaly of those modular flavor groups.
From Eqs.~(\ref{Z})-(\ref{ZT}) with $k/2=1/2$, (\ref{T8}), (\ref{M5gen}), and (\ref{M5alg}) as well as (\ref{M7gen}) and (\ref{M7alg}), we can obtain
\begin{align}
{\rm det}(a) = {\rm det}(b) = 1, \ {\rm det}(c) = {\rm det}(e^{\pi i/4}\mathbb{I}), \ {\rm det}(c)^8 =1. \label{detA5PSL2Z7Z8}
\end{align}
Actually, Eq.~(\ref{M5ST}) as well as Eq.~(\ref{M7ST}) satisfy Eq.~(\ref{detA5PSL2Z7Z8}).
Thus, in both cases, only $Z_8$ symmetry, generated by $c$, can be anomalous and then $A_5$ and $PSL(2,Z_7)$ remain anomaly-free.


\section{Modular symmetry on magnetized orbifolds of $T^2 \times T^2$}
\label{T2T2}

In this section, we extend the analyses to the modular symmetry on magnetized orbifolds of $T^2_1 \times T^2_2$, where both of the modulus on $T^2_i\ (i=1,2)$, $\tau_i$, are identified each other, i.e. $\tau_1 = \tau_2 \equiv \tau$. (See Ref.~\cite{Kikuchi:2020nxn}.)
First, let us consider the modular transformation for the wavefunctions on the $T^2_1/\mathbb{Z}_2^{({\rm t}_1)} \times T^2/\mathbb{Z}_2^{({\rm t}_2)}$ with the magnetic flux $M^{(i)}=$even and the SS phases $(\alpha_1^{(i)},\alpha_2^{(i)})=(0,0)$, and 
the magnetic flux $M^{(i)}=$odd and the SS phases $(\alpha_1^{(i)},\alpha_2^{(i)})=(1/2,1/2)$  on each $T^2_i/\mathbb{Z}_2^{({\rm t}_i)}$.
The wavefunctions transform under the modular transformation as
\begin{align}
&\Psi^{j^{(1)}j^{(2)},M^{(1)}M^{(2)}}_{({\rm t}_1)m_1({\rm t}_2)m_2}(\gamma(z_1,z_2,\tau)) \notag \\
&= {J}_1({\gamma},\tau) \sum_{k^{(1)}=0}^{N_{m_1}(M^{(1)})} \sum_{k^{(2)}=0}^{N_{m_2}(M^{(2)})} \rho_{({\rm t}_1)m_1({\rm t}_2)m_2}({\gamma})_{(j^{(1)}j^{(2)})(k^{(1)}k^{(2)})} \Psi^{k^{(1)}k^{(2)},M^{(1)}M^{(2)}}_{({\rm t}_1)m_1({\rm t}_2)m_2}(z_1,z_2,\tau) \label{modularformT2Z2T2Z2} \\
& \qquad m_1 \in \mathbb{Z}_2^{({\rm t}_1)}, \ m_2 \in \mathbb{Z}_2^{({\rm t}_2)}, \quad  \gamma \in \Gamma \notag \\
&\Psi^{j^{(1)}j^{(2)},M^{(1)}M^{(2)}}_{({\rm t}_1)m_1({\rm t}_2)m_2}(z_1,z_2,\tau) = \psi^{(k^{(1)}+\alpha_1^{(1)},\alpha_2^{(1)}),M^{(1)}}_{T^2_1/\mathbb{Z}_2^{({\rm t}_1)m_1}}(z_1,\tau) \psi^{(k^{(2)}+\alpha_1^{(2)},\alpha_2^{(2)}),M^{(2)}}_{T^2_2/\mathbb{Z}_2^{({\rm t}_2)m_2}}(z_2,\tau) \label{psiT2Z2T2Z2} \\
&\rho_{({\rm t}_1)m_1({\rm t}_2)m_2}({S})_{(j^{(1)}j^{(2)})(k^{(1)}k^{(2)})}
  = \widetilde{\rho}_{T^2_1/\mathbb{Z}_2^{({\rm t}_1)m_1}}(\widetilde{S})_{j^{(1)}k^{(1)}} \widetilde{\rho}_{T^2_2/\mathbb{Z}_2^{({\rm t}_2)m_2}}(\widetilde{S})_{j^{(2)}k^{(2)}}, \label{SonT2Z2T2Z2} \\
&\rho_{({\rm t}_1)m_1({\rm t}_2)m_2}({T})_{(j^{(1)}j^{(2)})(k^{(1)}k^{(2)})}
  = \widetilde{\rho}_{T^2_1/\mathbb{Z}_2^{({\rm t}_1)m_1}}(\widetilde{T})_{j^{(1)}k^{(1)}} \widetilde{\rho}_{T^2_2/\mathbb{Z}_2^{({\rm t}_2)m_2}}(\widetilde{T})_{j^{(2)}k^{(2)}},
\label{TonT2Z2T2Z2}
\end{align}
where $\widetilde{\rho}_{T^2_2/\mathbb{Z}_2^{({\rm t}_i)m_i}}(\widetilde{\gamma})\ (i=1,2)$ correspond to Eqs.~(\ref{rhoSandTZ2even})-(\ref{rhoSandTZ2odd}) for $M^{(i)}=$even and $(\alpha_1^{(i)},\alpha_2^{(i)})=(0,0)$ or Eqs.~(\ref{rhoSZ2evenSS12})-(\ref{rhoTZ2oddSS12}) for $M^{(i)}=$odd and $(\alpha_1^{(i)},\alpha_2^{(i)})=(1/2,1/2)$.
Then, $\rho_{({\rm t}_1)m_1({\rm t}_2)m_2}({\gamma})$ satisfies Eq.~(\ref{Gammarep}) with $k=1$, where $\rho(Z)=-\mathbb{I}$ is replaced by $\rho_{({\rm t}_1)m_1({\rm t}_2)m_2}(Z)=-(-1)^{m_1+m_2}\mathbb{I}$, and also satisfies\footnote{lcm$(a,b)$ denotes the least common multiple of $a$ and $b$, and gcd$(a,b)$ denotes the greatest common divisor of $a$ and $b$.}
\begin{align}
&\rho({T})^{2{\rm lcm}(M^{(1)},M^{(2)})} = \mathbb{I}, \quad (M^{(1)}=2s^{(1)}, M^{(2)}=2s^{(2)}), \label{2s2s} \\
&\rho({T})^{2{\rm lcm}(M^{(1)},M^{(2)})} = \mathbb{I}, \quad (M^{(1)}=4s^{(1)}, M^{(2)}=2s^{(2)}-1), \label{4s2s1} \\
&\rho({T})^{2{\rm lcm}(M^{(1)},M^{(2)})} = -\mathbb{I},\ \rho({T})^{4 {\rm lcm}(M^{(1)},M^{(2)})} = \mathbb{I}, \quad (M^{(1)}=2(2s^{(1)}-1), M^{(2)}=2s^{(2)}-1) \label{22s12s1} \\
&\rho({T})^{{\rm lcm}(M^{(1)},M^{(2)})} = e^{\pi i\frac{M^{(1)}+M^{(2)}}{4{\rm gcd}(M^{(1)},M^{(2)})}} \mathbb{I}, \quad (M^{(1)}=2s^{(1)}-1, M^{(2)}=2s^{(2)}-1) \notag \\
&\ \Rightarrow \rho({T})^N = \mathbb{I}, \quad N= \left\{
  \begin{array}{l}
    {\rm lcm}(M^{(1)},M^{(2)}) \quad (M^{(1)}+M^{(2)} \in 8\mathbb{Z}) \\
    2{\rm lcm}(M^{(1)},M^{(2)}) \quad (M^{(1)}+M^{(2)} \in 4\mathbb{Z}) \\
    4{\rm lcm}(M^{(1)},M^{(2)}) \quad (M^{(1)}+M^{(2)} \in 2\mathbb{Z}) \\
  \end{array}
  \right., \label{2s12s1}
\end{align}
corresponding to Eq.~(\ref{Gamma'Nrep}), where $s^{(1)}, s^{(2)} \in \mathbb{Z}$ and we omit the $\mathbb{Z}_2^{({\rm t})}$ indices since the above relations are independent of them.
Thus, the wavefunctions on the magnetized $T^2_1/\mathbb{Z}_2^{({\rm t}_1)} \times T^2/\mathbb{Z}_2^{({\rm t}_2)}$ orbifold behave as the modular forms of weight $1$  and then they transform as $N_{m_1}(M^{(1)})N_{m_2}(M^{(2)})$-dimensional representations, where $N_{m_i}(M^{(i)})\ (i=1,2)$ denote the number of zero-mode wavefunctions on $T^2_i/\mathbb{Z}_2^{({\rm t}_i)}$.
These can be irreducible representations.
We will study their flavor symmetries in the next section.
Also note that when $m_1+m_2=1$, $S^2=\mathbb{I}$ is satisfied even though the modular weight $k=1$.\footnote{This situation does not appear on modular forms and actually the wavefunctions vanish at $z_1=z_2=0$.}

We can further consider the $\mathbb{Z}_2^{\rm(p)}$ permutation orbifold if $M^{(1)}=M^{(2)}=M$, $\alpha_i^{(1)}=\alpha_i^{(2)}=\alpha_i\ (i=1,2)$, and $m_1=m_2=m$.
The $\mathbb{Z}_2^{\rm(p)}$ permutation means the transformation of the complex coordinate of $T^2_1/\mathbb{Z}_2^{({\rm t}_1)} \times T^2/\mathbb{Z}_2^{({\rm t}_2)}$: $(z_1,z_2) \rightarrow (z_2,z_1)$, and then the $\mathbb{Z}_2^{\rm(p)}$ permutation orbifold can be considered by identifying $z_1$ and $z_2$.
Hence, the wavefunctions on the $\mathbb{Z}_2^{\rm(p)}$ permutation orbifold of  $T^2_1/\mathbb{Z}_2^{({\rm t}_1)} \times T^2/\mathbb{Z}_2^{({\rm t}_2)}$, i.e. $(T^2_1 \times T^2_2)/(\mathbb{Z}_2^{\rm (t)}\times \mathbb{Z}_2^{\rm (p)})$ orbifold, are expressed as
\begin{align}
&\Psi^{j^{(1)}j^{(2)},M}_{(t)m(p)n}(z_1,z_2,\tau) = {\cal N}_{{\rm(t,p)}}^{j^{(1)}j^{(2)}} \left( \Psi^{j^{(1)}j^{(2)},MM}_{({\rm t})m({\rm t})m}(z_1,z_2,\tau) +(-1)^n \Psi^{j^{(1)}j^{(2)},MM}_{({\rm t})m({\rm t})m}(z_2,z_1,\tau) \right) \notag \\
&\quad m \in \mathbb{Z}_2^{{\rm(t)}}, \quad n \in \mathbb{Z}_2^{{\rm(p)}}, \quad j^{(1)} \geq j^{(2)}, \quad {\cal N}_{{\rm(t,p)}}^{j^{(1)}j^{(2)}} = \left\{
  \begin{array}{l}
    1/2 \quad (j^{(1)} = j^{(2)}) \\
    1/\sqrt{2} \quad (j^{(1)} > j^{(2)})
  \end{array}
  \right., \label{Z2tZ2pwave}
\end{align}
and they satisfy the following boundary condition,
\begin{align}
\Psi^{j^{(1)}j^{(2)},M}_{(t)m(p)n}(z_2,z_1,\tau) = (-1)^n \Psi^{j^{(1)}j^{(2)},M}_{(t)m(p)n}(z_1,z_2,\tau), \label{z1z2}
\end{align}
in addition to ones in Eqs.~(\ref{psiz1}), (\ref{psiztau}), and (\ref{-zT2Z2}).
Thus, we can obtain $N_m(M)(N_m(M)+1)/2$-number of $\mathbb{Z}_2^{{\rm(p)}}$-even ($n=0$) modes and $N_m(M)(N_m(M)-1)/2$-number of $\mathbb{Z}_2^{{\rm(p)}}$-odd ($n=1$) modes.
We show the number of ($\mathbb{Z}_2^{\rm (t)}$ twist, $\mathbb{Z}_2^{\rm (p)}$ permutation)-eigenmodes, $N_{(m,n)}(M)=N_m(M)(N_m(M)+(-1)^n)/2$, which have the modular symmetry in Tables \ref{Z2tZ2p} and \ref{Z2tZ2pSS}.
Under the modular transformation, the wavefunctions in Eq.~(\ref{Z2tZ2pwave}) similarly transform as Eq.~(\ref{modularformT2Z2T2Z2}) replacing Eqs.~(\ref{SonT2Z2T2Z2}) and (\ref{TonT2Z2T2Z2}) with
\begin{align}
  &\rho_{({\rm t})m({\rm p})n}({\gamma})_{(j^{(1)}j^{(2)})(k^{(1)}k^{(2)})} \notag \\
  &= 2{\cal N}_{{\rm(t,p)}}^{j^{(1)}j^{(2)}} {\cal N}_{{\rm(t,p)}}^{k^{(1)}k^{(2)}} \left( \rho_{({\rm t})m}({\gamma})_{(j^{(1)}j^{(2)})(k^{(1)}k^{(2)})} + (-1)^n \rho_{({\rm t})m}({S})_{(j^{(1)}j^{(2)})(k^{(1)}k^{(2)})} \right), \label{UniReponPerm}
\end{align}
where it satisfies Eq.~(\ref{Gammarep}) with $k=1$ and also satisfies
\begin{align}
\rho_{({\rm t})m({\rm p})n}({T})^{2M} &= \mathbb{I}, \quad &(M \in 2\mathbb{Z}), \label{T2M} \\
\rho_{({\rm t})m({\rm p})n}({T})^M = i\mathbb{I}, \quad \rho_{({\rm t})m({\rm p})n}({T})^{2M} &= -\mathbb{I}, \quad \rho_{({\rm t})m({\rm p})n}({T})^{4M} = \mathbb{I}, \quad  &(M \in 2\mathbb{Z}+1), \label{T4M}
\end{align}
corresponding to Eq.~(\ref{Gamma'Nrep}).
Thus, the wavefunctions on the $(T^2_1 \times T^2_2)/(\mathbb{Z}_2^{\rm (t)}\times \mathbb{Z}_2^{\rm (p)})$ orbifold with the magnetic flux $M \in 2\mathbb{Z} $ and the SS phases $(\alpha_1,\alpha_2)=(0,0) $ behave as the modular forms of weight $1$ and then they transform as $N_{(m,n)}(M)$-dimensional representations, as shown in Table \ref{Z2tZ2p}.
Similarly, the wavefunctions with the magnetic flux $M \in 2\mathbb{Z}+1$ and the SS phases $(\alpha_1,\alpha_2)=(1/2,1/2)$ also behave as the modular forms of weight $1$ and then they transform as $N_{(m,n)}(M)$-dimensional representations, as shown in Table \ref{Z2tZ2pSS}.

\begin{table}[H]
  \centering
  \begin{tabular}{|c|c|c|c|c|c|} \hline
     & $M$ & 2 & 4 & 6 & 8 \\ \hline
    {\rm (even,\ even)}: $N_{(0,0)}(M)$ & $(M+2)(M+4)/8$ & \fbox{3} & 6 & 10 & 15 \\ \hline
    {\rm (even,\ odd)}: $N_{(0,1)}(M)$ & $M(M+2)/8$ & 1 & \fbox{3} & 6 & 10 \\ \hline
    {\rm (odd,\ even)}: $N_{(1,0)}(M)$ & $M(M-2)/8$ & 0 & 1 & \fbox{3} & 6 \\ \hline
    {\rm (odd,\ odd)}: $N_{(1,1)}(M)$ & $(M-2)(M-4)/8$ & 0 & 0 & 1 & \fbox{3} \\ \hline
    order $h$ of $T$ ($T^h=\mathbb{I}$) & ${2M}$ & $4$ & ${8}$ & ${12}$ & ${16}$ \\ \hline
\end{tabular}
\caption{The number of ($\mathbb{Z}_2^{\rm (t)}$ twist, $\mathbb{Z}_2^{\rm (p)}$ permutation)-eigenmodes, $N_{(m,n)}(M)$, on the $(T^2_1 \times T^2_2)/(\mathbb{Z}_2^{\rm (t)}\times \mathbb{Z}_2^{\rm (p)})$ orbifold with $M=$even and $(\alpha_1,\alpha_2)=(0,0)$, and the order of $T$. The three generations are boxed.}
\label{Z2tZ2p}
  \centering
  \begin{tabular}{|c|c|c|c|c|c|} \hline
     & $M$ & 1 & 3 & 5 & 7 \\ \hline
    {\rm (even,\ even)}: $N_{(0,0)}(M)$ & $(M-1)(M+1)/8$ & 0 & 1 & \fbox{3} & 6 \\ \hline
    {\rm (even,\ odd)}: $N_{(0,1)}(M)$ & $(M-1)(M-3)/8$ & 0 & 0 & 1 & \fbox{3} \\ \hline
    {\rm (odd,\ even)}: $N_{(1,0)}(M)$ & $(M+1)(M+3)/8$ & 1 & \fbox{3} & 6 & 10 \\ \hline
    {\rm (odd,\ odd)}: $N_{(1,1)}(M)$ & $(M+1)(M-1)/8$ & 0 & 1 & \fbox{3} & 6 \\ \hline
   order $h$ of $T$  ($T^h=\mathbb{I}$) & ${4M}$ & $4$ & ${12}$ & ${20}$ & ${28}$ \\ \hline
\end{tabular}
\caption{The number of ($\mathbb{Z}_2^{\rm (t)}$ twist, $\mathbb{Z}_2^{\rm (p)}$ permutation)-eigenmodes, $N_{(m,n)}(M)$, on the $(T^2_1 \times T^2_2)/(\mathbb{Z}_2^{\rm (t)}\times \mathbb{Z}_2^{\rm (p)})$ orbifold with $M=$odd and $(\alpha_1,\alpha_2)=(1/2,1/2)$, and the order of $T$. The three generations are boxed.}
\label{Z2tZ2pSS}
\end{table}

In the next section, we show the specific modular flavor groups of the three-generation modes on the magnetized orbifolds of $T^2 \times T^2$.


\section{Modular flavor groups of three-generation modes on magnetized orbifolds of $T^2 \times T^2$}
\label{T2T2three}

\subsection{$(T^2_1 \times T^2_2)/(\mathbb{Z}_2^{\rm (t)}\times \mathbb{Z}_2^{\rm (p)})$ orbifold}

Firstly, we consider the three-generation modes on the magnetized $(T^2_1 \times T^2_2)/(\mathbb{Z}_2^{\rm (t)}\times \mathbb{Z}_2^{\rm (p)})$ orbifold in Tables \ref{Z2tZ2p} and \ref{Z2tZ2pSS}.

As shown in Table \ref{Z2tZ2p}, we can obtain four models with three-generation modes on the $(T^2_1 \times T^2_2)/(\mathbb{Z}_2^{\rm (t)}\times \mathbb{Z}_2^{\rm (p)})$ orbifold with $M=$even and $(\alpha_1,\alpha_2)=(0,0)$: $(M,m,n)=(2,0,0)$, $(4,0,1)$, $(6,1,0)$, and $(8,1,1)$.
They can be repesentations of $\Delta'(6M^2)$, which are the double covering groups of $\Delta(6M^2)$, similar as shown in subsection \ref{T2even}.
Namely, if Eq.~(\ref{X}) is also satisfied\footnote{When $M=1,2$, Eq.~(\ref{X}) is automatically satisfied by considering Eq.~(\ref{Gammarep}). (See in detail Appendix~\ref{proof}.)} in addition to Eqs.~(\ref{Gammarep}) and (\ref{Gamma'Nrep}) with $k=1$ and $N=2M$, the following generators,
\begin{align}
&a=ST^2ST^4, \ a'=ST^2S^{-1}T^{-2}, \ b=T^{\frac{M}{2}+3}S^{\frac{3}{2}M-1}T^{M}, \ c=ST^{M-2}ST^{\frac{3}{2}M-1},\ (M=4s) \label{Delta'ST} \\
&a=ST^2ST^4, \ a'=ST^2S^{-1}T^{-2}, \ b=T^{\frac{M}{2}}S^{\frac{3}{2}M}T^{M}, \quad c=ST^{M}ST^{\frac{3}{2}M},\ (M=2(2s-1)) \label{Delta'STodd}
\end{align}
where $s \in \mathbb{Z}$, satisfy
\begin{align}
&a^M=a'^M=b^3=c^4=\mathbf{1}, \label{algDelta'} \\
&aa'=a'a, \ cbc^{-1}=b^{-1}, \ bab^{-1}=a^{-1}a'^{-1}, \ ba'b^{-1}=a, \ cac^{-1}=a'^{-1}, ca'c^{-1}=a^{-1}, \notag
\end{align}
which means the generators in Eq.~(\ref{Delta'ST}) are ones of $\Delta'(6M^2) \simeq (Z_M \times Z_M) \rtimes Z_3 \rtimes Z_4 \simeq \Delta(3M^2) \rtimes Z_4$, where $a^{(')}$, $b$, $c$ denote ones of $Z^{(')}_M$, $Z_3$, $Z_4$, respectively.
Actually, all of the following $S$ and $T$ transformation matrices for $(M,m,n)=(2,0,0)$, $(4,0,1)$, $(6,1,0)$, and $(8,1,1)$ satisfy Eq.~(\ref{X}) since they form as Eq.~(\ref{STingen}).

The $S$ and $T$ transformation matrices for $(M,m,n)=(2,0,0)$ are given by
\begin{align}
S = \frac{i}{2}
\begin{pmatrix}
1 & \sqrt{2} & 1 \\
\sqrt{2} & 0 & -\sqrt{2} \\
1 & -\sqrt{2} & 1
\end{pmatrix},
\quad&
T = 
\begin{pmatrix}
1 & \ & \ \\
\ & i & \ \\
\ & \ & -1
\end{pmatrix}.
\label{M2M2ST}
\end{align}

The $S$ and $T$ transformation matrices for $(M,m,n)=(4,0,1)$ are given by
\begin{align}
S = - \frac{i}{2}
\begin{pmatrix}
1 & \sqrt{2} & 1 \\
\sqrt{2} & 0 & -\sqrt{2} \\
1 & -\sqrt{2} & 1
\end{pmatrix},
\quad&
T = e^{\pi i/4}
\begin{pmatrix}
1 & \ & \ \\
\ & e^{3\pi i/4} & \ \\
\ & \ & -1
\end{pmatrix}.
\label{M4M4ST}
\end{align}

The $S$ and $T$ transformation matrices for $(M,m,n)=(6,1,0)$ are given by
\begin{align}
S = - \frac{i}{2}
\begin{pmatrix}
1 & \sqrt{2} & 1 \\
\sqrt{2} & 0 & -\sqrt{2} \\
1 & -\sqrt{2} & 1
\end{pmatrix},
\quad&
T = e^{\pi i/3}
\begin{pmatrix}
1 & \ & \ \\
\ & i & \ \\
\ & \ & -1
\end{pmatrix}.
\label{M6M6ST}
\end{align}

The $S$ and $T$ transformation matrices for $(M,m,n)=(6,1,0)$ are given by
\begin{align}
S = \frac{i}{2}
\begin{pmatrix}
1 & \sqrt{2} & 1 \\
\sqrt{2} & 0 & -\sqrt{2} \\
1 & -\sqrt{2} & 1
\end{pmatrix},
\quad&
T = e^{5\pi i/8}
\begin{pmatrix}
1 & \ & \ \\
\ & e^{5\pi i/8} & \ \\
\ & \ & -1
\end{pmatrix}.
\label{M8M8ST}
\end{align}
Note that since the $T$ matrix in Eq.~(\ref{M6M6ST}) also satisfies $T^4=e^{4\pi i/3}\mathbb{I}$, this can be the $Z_3$ generator, $d=T^4$, which commutes with all the generators in Eq.~(\ref{Delta'STodd}) and also the generators $a$ and $a'$ in Eq.~(\ref{Delta'STodd}) satisfy $a^{2}=a'^{2}=\mathbf{1}$.
Thus, the three-generation modes for $(M,m,n)=(2,0,0)$, $(4,0,1)$, $(6,1,0)$, and $(8,1,1)$ are transformed under the modular transformation as the three-dimensional representations of $S'_4 \simeq \Delta'(24)$, $\Delta'(96)$, $S'_4 \times Z_3$, and $\Delta'(384)$, respectively.

We also comment on the anomaly of those modular flavor groups.
From Eqs.~(\ref{Gammarep}) with $k=1$, (\ref{Gamma'Nrep}) with $N=2M$, (\ref{Delta'ST}), (\ref{Delta'STodd}), and (\ref{algDelta'}), similarly, we can obtain
\begin{align}
{\rm det}(a) = {\rm det}(a') = {\rm det}(b) = 1, \ 
{\rm det}(c) = \left\{
  \begin{array}{l}
    {\rm det}(T)^{\frac{M}{2}+3} \ (M=4s) \\
    {\rm det}(T)^{\frac{M}{2}+6} \ (M=2(2s-1))
  \end{array}
  \right., \ {\rm det}(c)^4 =1. \label{detDelta'}
\end{align}
All of Eqs.~(\ref{M2M2ST})-(\ref{M8M8ST}) satisfy Eq.~(\ref{detDelta'}) and  ${\rm det}(c)=i$.
In Eq.~(\ref{M6M6ST}), ${\rm det}(d)={\rm det}(T)^4=1$ is also satisfied.
Thus, in all cases, only $Z_4$ symmetry, generated by $c$, can be  anomalous and then $A_4 \simeq \Delta(12)$, $\Delta(48)$, $A_4 \times Z_3$, and $\Delta(192)$ remain anomaly-free.

As shown in Table \ref{Z2tZ2pSS}, we can obtain four models with three-generation modes on the $(T^2_1 \times T^2_2)/(\mathbb{Z}_2^{\rm (t)}\times \mathbb{Z}_2^{\rm (p)})$ orbifold with $M=$odd and $(\alpha_1,\alpha_2)=(1/2,1/2)$: $(M,m,n)=(3,1,0)$, $(5,0,0)$, $(5,1,1)$, and $(7,0,1)$.
We note that all of the following $S$ and $T$ transformation matrices satisfy Eqs.~(\ref{Gammarep}) and (\ref{T4M}) with $k=1$.
First, from the following $S$ and $T$ transformation matrices for $(M,m,n)=(3,1,0)$,
\begin{align}
S = - \frac{i}{3}
\begin{pmatrix}
1 & 2 e^{\pi i/3} & 2 e^{2\pi i/3} \\
2 e^{-\pi i/3} & 1 & - 2 e^{\pi i/3} \\
2 e^{-2\pi i/3} & - 2 e^{-\pi i/3} & 1
\end{pmatrix}, \ 
T = 
\begin{pmatrix}
e^{\pi i/6} & \ & \ \\
\ & e^{5\pi i/6} & \ \\
\ & \ & e^{9\pi i/6}
\end{pmatrix},
\label{M3M3ST}
\end{align}
we can obtain the following generators,
\begin{align}
a=ST^9, \ b=ST, \ c=T^3 \label{M3M3gen}
\end{align}
satisfying
\begin{align}
a^2=b^3=(ab)^3=c^4=\mathbf{1}, \ ac=ca, \ bc=cb \label{M3M3alg}
\end{align}
which mean the generators in Eq.~(\ref{M3M3gen}) are ones of $A_4 \times Z_4$.
Thus, the three-generation modes, $(M,m,n)=(3,1,0)$, are transformed under the modular transformation as the three-dimensional representations of $A_4 \times Z_4$.

Second, from the following $S$ and $T$ transformation matrices for $(M,m,n)=(5,0,0)$,
\begin{align}
S &= \frac{4i}{5}
\begin{pmatrix}
A^2 & \sqrt{2} e^{\pi i/5} AB & e^{2\pi i/5}B^2 \\
\sqrt{2} e^{-\pi i/5} AB & B^2-A^2 & - \sqrt{2} e^{\pi i/5} AB \\
e^{-2\pi i/5}B^2 & - \sqrt{2} e^{-\pi i/5} AB & A^2
\end{pmatrix}, \ 
T &= 
\begin{pmatrix}
e^{\pi i/10} & \ & \ \\
\ & e^{5\pi i/10} & \ \\
\ & \ & e^{9\pi i/10}
\end{pmatrix},
\label{M5M5eST} \\
&\qquad A = \cos \left(\frac{\pi}{10}\right), \ B = \cos \left(\frac{3\pi}{10}\right), \notag
\end{align}
we can obtain the following generators,
\begin{align}
a=ST^5, \ b=ST, \ c=T^5 \label{M5M5gen}
\end{align}
satisfying
\begin{align}
a^2=b^3=(ab)^5=c^4=\mathbf{1}, \ ac=ca, \ bc=cb \label{M5M5alg}
\end{align}
which mean the generators in Eq.~(\ref{M5M5gen}) are ones of $A_5 \times Z_4$.
Thus, the three-generation modes, $(M,m,n)=(5,0,0)$, are transformed under the modular transformation as the three-dimensional representations of $A_5 \times Z_4$.

Third, similarly, from the following $S$ and $T$ transformation matrices for $(M,m,n)=(5,1,1)$,
\begin{align}
S &= - \frac{2i}{5}
\begin{pmatrix}
2 \left(A^2 - B^2\right) & - \sqrt{2} e^{\pi i/5} \left(A+B\right) & - \sqrt{2} e^{2\pi i/5} \left(A+B\right) \\
- \sqrt{2} e^{-\pi i/5} \left(A+B\right) & A- 1 & e^{\pi i/5} \left(B + 1\right) \\
- \sqrt{2} e^{-2\pi i/5} \left(A+B\right) & e^{-\pi i/5} \left(B + 1\right) & A-1
\end{pmatrix},
\notag \\
&\qquad A = \sin \left(\frac{\pi}{10}\right), \ B = \sin \left(\frac{3\pi}{10}\right), \notag \\
T &= 
\begin{pmatrix}
e^{5\pi i/10} & \ & \ \\
\ & e^{13\pi i/10} & \ \\
\ & \ & e^{17\pi i/10}
\end{pmatrix}
\label{M5M5oST}
\end{align}
we can obtain the generators in Eq.~(\ref{M5M5gen}) satisfying Eq.~(\ref{M5M5alg}).
Thus, the three-generation modes, $(M,m,n)=(5,1,1)$, are also transformed under the modular transformation as the three-dimensional representations of $A_5 \times Z_4$.

Fourth, from the following $S$ and $T$ transformation matrices for $(M,m,n)=(7,0,1)$,
\begin{align}
&S = \frac{4i}{7}
\begin{pmatrix}
AD-B^2 & - e^{\frac{\pi i}{7}} \left(A^2+BC\right) & - e^{\frac{2\pi i}{7}} \left(AB+CD\right) \\
- e^{-\frac{\pi i}{7}} \left(A^2+BC\right) & AB-C^2 & e^{\frac{\pi i}{7}} \left(B^2+AC\right) \\
- e^{-\frac{2\pi i}{7}} \left(AB+CD\right) & e^{-\frac{\pi i}{7}} \left(B^2+AC\right) & BD-A^2
\end{pmatrix}, \notag \\
&\qquad A = \cos \left(\frac{\pi}{14}\right), \ B = \cos \left(\frac{3\pi}{14}\right), \ C = \cos \left(\frac{5\pi}{14}\right), \ D = \cos \left(\frac{9\pi}{14}\right), \notag \\
&T = 
\begin{pmatrix}
e^{5\pi i/14} & \ & \ \\
\ & e^{13\pi i/14} & \ \\
\ & \ & e^{17\pi i/14}
\end{pmatrix},
\label{M7M7ST}
\end{align}
which also satisfy Eq.~(\ref{add}), we can obtain the following generators,
\begin{align}
a=ST^{21}, \ b=S^3T^3, \ c=T^7 \label{M7M7gen}
\end{align}
satisfying
\begin{align}
a^2=b^4=(ab)^7=(a^{-1}b^{-1}ab)^4=c^4=\mathbf{1}, \ ac=ca, \ bc=cb \label{M7M7alg}
\end{align}
which mean the generators in Eq.~(\ref{M7M7gen}) are ones of $PSL(2,Z_7) \times Z_4$.
Thus, the three-generation modes, $(M,m,n)=(7,0,1)$, are transformed under the modular transformation as the three-dimensional representations of $PSL(2,Z_7) \times Z_4$.

Finally, we also comment on the anomaly of those modular flavor groups.
From Eqs.~(\ref{M3M3ST})-(\ref{M7M7gen}), and also Eqs.~(\ref{Gammarep}) and (\ref{T4M}) with $k=1$, we can obtain
\begin{align}
{\rm det}(a) = {\rm det}(b) = 1, \ {\rm det}(c) = {\rm det}(i\mathbb{I})=-i, \ {\rm det}(c)^4 =1. \label{detModdModd}
\end{align}
Thus, in all the above cases, only $Z_4$ symmetry, generated by $c$, can be anomalous and then $A_4$, $A_5$ and $PSL(2,Z_7)$ remain anomaly-free.

Therefore, on the magnetized $(T^2_1 \times T^2_2)/(\mathbb{Z}_2^{\rm (t)}\times \mathbb{Z}_2^{\rm (p)})$ orbifold, we can obtain three-dimensional representations of all the double covering groups of $\Gamma_4 \simeq S_4$,  $\Gamma_8 \supset \Delta(96)$, and $\Gamma_{16} \supset \Delta(384)$ for even magnetic fluxes and 
$Z_4$ central extended groups of  $\Gamma_3 \simeq PSL(2,\mathbb{Z}_3)\simeq A_4$, $\Gamma_5 \simeq PSL(2,\mathbb{Z}_5)\simeq A_5$, $\Gamma_7\simeq PSL(2,\mathbb{Z}_7)$ for odd magnetic fluxes.


\subsection{Other  $T^2_1/\mathbb{Z}_2^{({\rm t}_1)} \times T^2/\mathbb{Z}_2^{({\rm t}_2)}$ orbifold}

Finally, we consider the three-generation modes on the magnetized $T^2_1/\mathbb{Z}_2^{({\rm t}_1)} \times T^2/\mathbb{Z}_2^{({\rm t}_2)}$ orbifold, where $T^2_1/\mathbb{Z}_2^{({\rm t}_1)}$ and  $T^2/\mathbb{Z}_2^{({\rm t}_2)}$ are not identified.
In order to obtain $N_{m_1}(M^{(1)})N_{m_2}(M^{(2)})=3$ on the magnetized $T^2_1/\mathbb{Z}_2^{({\rm t}_1)} \times T^2/\mathbb{Z}_2^{({\rm t}_2)}$ orbifold, we can only consider $N_{m_1}(M^{(1)})=3$ and $N_{m_2}(M^{(2)})=1$.
Then, from Tables \ref{even} and \ref{odd}, we can consider twelve patterns, listed in Table \ref{Z2t3Z2t1}.
The corresponding finite modular subgroups which can be found by considering $Z=-(-1)^{m_1+m_2}\mathbf{1}$ and  Eqs.~(\ref{2s2s})-(\ref{2s12s1})~\footnote{There is an exception in Eq.~(\ref{4s2s1}); Eq.~(\ref{4s2s1}) for $M^{(1)}=4$ singlet mode, that is $\mathbb{Z}_2^{({\rm t}_1)}$-odd mode of $M^{(1)}=4$, corresponds to Eq.~(\ref{2s12s1}) with $M^{(1)}=1$.} are also listed in Table \ref{Z2t3Z2t1}.
The $S$ and $T$ transformation matrices for the $\mathbb{Z}_2^{{\rm(t)}}$-odd modes with $M=1$ as well as the $\mathbb{Z}_2^{{\rm(t)}}$-odd modes with $M=4$ are given by
\begin{align}
S = e^{3\pi i/4}, \ T = e^{\pi i/4},
\end{align}
and ones for the $\mathbb{Z}_2^{{\rm(t)}}$-even modes with $M=3$ are given by
\begin{align}
S = e^{\pi i/4}, \ T = e^{\pi i/12},
\end{align}
while ones for $N_{m_1}(M^{(1)})=3$ modes are expressed in section \ref{T2three}.
Then, we can find the specific modular flavor groups as shown in Table \ref{Z2t3Z2t1}.
We also show the anomaly-free groups of them in Table \ref{Z2t3Z2t1}.

\begin{table}[H]
\centering
\begin{tabular}{|c|c|c|c|c|} \hline
$(M^{(1)},m_1:M^{(2)},m_2)$ & orders $(h_S,h_T)$ of  $S$ and $T$ & modular flavor group & anomaly-free group \\
& $(S^{h_S}=T^{h_T}=\mathbb{I})$ & &   \\ \hline
$(4,0:4,1)$ & $(2,8)$ & $\Delta(96)$ & $\Delta(96)$ \\ \hline
$(4,0:1,1)$ & $(2,8)$ & $\Delta(96)$ & $\Delta(96)$ \\ \hline
$(4,0:3,0)$ & $(4,24)$ & $\Delta'(96) \times Z_3$ & $\Delta(48) \times Z_3$ \\ \hline
$(8,1:4,1)$ & $(4,16)$ & $\Delta'(384)$ & $\Delta(192)$ \\ \hline
$(8,1:1,1)$ & $(4,16)$ & $\Delta'(384)$ & $\Delta(192)$ \\ \hline
$(8,1:3,0)$ & $(2,48)$ & $\Delta(384) \times Z_3$ & $\Delta(384) \times Z_3$ \\ \hline
$(5,1:4,1)$ & $(4,20)$ & $A_5 \times Z_4$ & $A_5$ \\ \hline
$(5,1:1,1)$ & $(4,20)$ & $A_5 \times Z_4$ & $A_5$ \\ \hline
$(5,1:3,0)$ & $(2,15)$ & $A_5 \times Z_3$ & $A_5 \times Z_3$ \\ \hline
$(7,0:4,1)$ & $(2,7)$ & $PSL(2,Z_7)$ & $PSL(2,Z_7)$ \\ \hline
$(7,0:1,1)$ & $(2,7)$ & $PSL(2,Z_7)$ & $PSL(2,Z_7)$ \\ \hline
$(7,0:3,0)$ & $(4,84)$ & $PSL(2,Z_7) \times Z_3 \times Z_4$ & $PSL(2,Z_7) \times Z_3$ \\ \hline
\end{tabular}
\caption{The flavor groups of the three-generation modes, $(M^{(1)},m_1:M^{(2)},m_2)$, which satisfy $N_{m_1}(M^{(1)})=3$ and $N_{m_2}(M^{(2)})=1$, on the magnetized $T^2_1/\mathbb{Z}_2^{({\rm t}_1)} \times T^2/\mathbb{Z}_2^{({\rm t}_2)}$ orbifold.
The anomaly-free subgroups are also shown.
}
\label{Z2t3Z2t1}
\end{table}


\section{Conclusion}
\label{conclusion}

We have studied the modular symmetry of wavefunctions on magnetized orbifolds: $T^2/\mathbb{Z}_2^{{\rm(t)}}$ twisted orbifold, $T^2_1/\mathbb{Z}_2^{({\rm t}_1)} \times T^2/\mathbb{Z}_2^{({\rm t}_2)}$ twisted orbifold, and the $\mathbb{Z}_2^{\rm(p)}$ permutation orbifold, i.e.~ $(T^2_1 \times T^2_2)/(\mathbb{Z}_2^{\rm (t)}\times \mathbb{Z}_2^{\rm (p)})$ orbifold, with the Scherk-Schwarz phases.
It has been found that we can consider the modular symmetry of not only wavefunctions with the magnetic flux $M=$even and the vanishing SS phases $(\alpha_1,\alpha_2)=(0,0)$ but also ones with the magnetic flux $M=$odd and the SS phases $(\alpha_1,\alpha_2)=(1/2,1/2)$.

Moreover, we have investigated the specific modular flavor groups for three-generation modes on the magnetized orbifolds.
The three-generation modes on the magnetized $T^2/\mathbb{Z}_2^{{\rm(t)}}$ twisted orbifold with the magnetic flux $M=4,8$ are three-dimensional representations of $\widetilde{\Delta}(96)$, $\widetilde{\Delta}(384)$, which are quadruple covering groups of $\Delta(96)$, $\Delta(384)$, respectively. Among them, only $Z_8$ symmetries can be anomalous and then $\Delta(48)$, $\Delta(192)$ are anomaly free, respectively. Note that since the anomalous $Z_8$ symmetry is discrete subgroup of $U(1)$, it can be canceled by the Green-Schwarz mechanism.
The three-generation modes on the magnetized $T^2/\mathbb{Z}_2^{{\rm(t)}}$ twisted orbifold with the magnetic flux $M=5,7$ are three-dimensional representations of $A_5 \times Z_8$, $PSL(2,Z_7) \times Z_8$, respectively. Among them, only $Z_8$ symmetries can  be  anomalous and then $A_5$ and $PSL(2,Z_7)$ are anomaly free, respectively.
Similarly, the three-generation modes on the magnetized $(T^2_1 \times T^2_2)/(\mathbb{Z}_2^{\rm (t)}\times \mathbb{Z}_2^{\rm (p)})$ orbifold are the corresponding three-dimensional representations of the double covering groups of $\Gamma_N$ for 
$N=4,8,16$ and $Z_4$ central extended groups of $\Gamma_N$ for $N=3,5,7$, provided in Ref.~\cite{deAdelhartToorop:2011re}. Among them, only $Z_8$ symmetries can be anomalous and then $\Delta(3M^2)$ for $N=2M=4,8,16$, $A_4$ for $N=3$, $A_5$ for $N=5$, $PSL(2,Z_7)$ for $N=7$ are anomaly free.
We have also showed the specific modular flavor groups of the three-generation modes on the other distinguishable magnetized  $T^2_1/\mathbb{Z}_2^{({\rm t}_1)} \times T^2/\mathbb{Z}_2^{({\rm t}_2)}$ orbifolds in Table \ref{Z2t3Z2t1}.

Our results on flavor symmetries of three generations 
are useful to understand quarks and lepton masses and their mixing 
angles.
Also, anomaly behaviors are useful. (See e.g. \cite{Kobayashi:2019mna}.)
We would  investigate the realistic model building considering the obtained modular flavor groups in magnetized orbifold models elsewhere.

%

\vspace{1.5 cm}
\noindent
{\large\bf Acknowledgement}\\

T. K. was supported in part by MEXT KAKENHI Grant Number JP19H04605. 
H. U. was supported by Grant-in-Aid for JSPS Research Fellows No. 20J20388.


\appendix


\section{Scherk-Schwarz phases and Wilson lines}
\label{SSWL}

Here, we show that the Scherk-Schwarz (SS) phases can be converted into the Wilison lines (WLs) through gauge transformation~\cite{Abe:2013bca} and also that the modular transformations for them are consistent each other.

First, let us consider the following gauge transformation,
\begin{align}
\widetilde{\psi}^{\alpha_1,\alpha_2}(z,\tau) &= e^{-i{\rm Re}\bar{\beta}z} \psi^{\alpha_1,\alpha_2}(z,\tau), \\
\widetilde{A}(z) = A(z) - d[{\rm Re}\bar{\beta}z] &= \frac{\pi M}{{\rm Im}\tau} {\rm Im}\left( \left(\bar{z} - \frac{i{\rm Im}\tau}{\pi M}\bar{\beta} \right) dz \right),
\end{align}
where $\beta$ is a complex number, $\psi^{\alpha_1,\alpha_2}$ satisfies Eqs.~(\ref{psiz1SS}) and (\ref{psiztauSS}), $A(z)$ is in Eq.~(\ref{A}).
We can regard $\frac{i{\rm Im}\tau}{\pi M}\beta \equiv \widetilde{a}_w$ as the WL.
Accordingly, $\chi_1(z)$ and $\chi_2(z)$, defined in Eqs.~(\ref{chi1}) and (\ref{chi2}), are deformed as
\begin{align}
\widetilde{\chi}_1(z) &= \frac{\pi M}{{\rm Im}\tau} {\rm Im}\left(z+\frac{i{\rm Im}\tau}{\pi M}\beta\right) = \chi_1(z) + {\rm Re}\beta, \\
\widetilde{\chi}_2(z) &= \frac{\pi M}{{\rm Im}\tau} {\rm Im}\bar{\tau}\left(z+\frac{i{\rm Im}\tau}{\pi M}\beta\right) = \chi_2(z) + {\rm Re}\bar{\tau}\beta.
\end{align}
Therefore, the boundary conditions of the gauge transformed wavefunction $\widetilde{\psi}^{\alpha_1,\alpha_2}$ are modified from Eqs.~(\ref{psiz1SS}) and (\ref{psiztauSS}) as
\begin{align}
&\widetilde{\psi}^{\alpha_1,\alpha_2}(z+1,\tau) = e^{2\pi i \alpha_1-2i{\rm Re}\beta}e^{i\widetilde{\chi}_1(z)} \widetilde{\psi}^{\alpha_1,\alpha_2}(z,\tau), \label{psiz1SSmodi} \\
&\widetilde{\psi}^{\alpha_1,\alpha_2}(z+\tau,\tau) = e^{2\pi i \alpha_2-2i{\rm Re}\bar{\tau}\beta}e^{i\widetilde{\chi}_2(z)} \widetilde{\psi}^{\alpha_1,\alpha_2}(z,\tau). \label{psiztauSSmodi}
\end{align}
When we chose $\beta = -i\pi \frac{\alpha_1\tau-\alpha_2}{{\rm Im}\tau}$, the gauge transformed wavefunction,
\begin{align}
\widetilde{\psi}^{\alpha_1,\alpha_2}(z,\tau) &= e^{\pi i\frac{{\rm Im}(\alpha_1\bar{\tau}-\alpha_2)z}{{\rm Im}\tau}} \psi^{\alpha_1,\alpha_2}(z,\tau), 
\label{SStoWL}
\end{align}
has the WL, $M\widetilde{a}_w=\alpha_1\tau-\alpha_2$, and the vanishing SS phases, $(\widetilde{\alpha}_1,\widetilde{\alpha}_2)=(0,0)$.
That is, the SS phases, $(\alpha_1,\alpha_2)$, can be converted into the WL, $M\widetilde{a}_w=\alpha_1\tau-\alpha_2$, through gauge transformation in Eq.~(\ref{SStoWL}).
Actually, the $j$-th wavefunction can be expressed as
\begin{align}
\widetilde{\psi}_{T^2}^{(j+\alpha_1,\alpha_2),M}(z,\tau) = e^{-\pi i\frac{\alpha_1\alpha_2}{M}} \psi_{T^2}^{(j+0,0),M}(z+\widetilde{a}_w,\tau). \label{psiWL}
\end{align}

Next, let us consider the modular transformation.
When $M=$even ($x=0$), the WL transforms as
\begin{align}
&T(M\widetilde{a}_w) = \alpha_1(\tau+1) - (\alpha_1+\alpha_2) = M\widetilde{a}_w, \quad
T=
\begin{pmatrix}
1 & 1 \\
0 & 1
\end{pmatrix}, \\
&S(M\widetilde{a}_w) = -\alpha_2\left(-\frac{1}{\tau}\right) - \alpha_1 = \frac{M\widetilde{a}_w}{-\tau}, \qquad
S=
\begin{pmatrix}
0 & 1 \\
-1 & 0
\end{pmatrix},
\end{align}
that is, it transforms as
\begin{align}
&\gamma(M\widetilde{a}_w) = \frac{M\widetilde{a}_w}{c\tau+d}, \quad
\gamma=
\begin{pmatrix}
a & b \\
c & d
\end{pmatrix}.
\end{align}
In this case, as mentioned in Ref.~\cite{Kikuchi:2020frp}, the modular transformation for the wavefunction in right-hand side of Eq.~(\ref{psiWL}) is the same as Eq.~(\ref{rhoSandT}).
Furthermore, in this case, the gauge phase in Eq.~(\ref{SStoWL}) is invariant under the modular transformation and then the modular transformation for the gauge transformed wavefunction in left hand side of Eq.~(\ref{SStoWL}) or Eq.~(\ref{psiWL}) is the same as Eqs.~(\ref{rhoSSS}) and (\ref{rhoTSS}).
These are consistent.
When $M=$odd ($x=1$), the $T$ transformation for the WL as
\begin{align}
T(M\widetilde{a}_w) = \alpha_1(\tau+1) - \left(\alpha_1+\alpha_2-\frac{M}{2}\right) = M\left(\widetilde{a}_w+\frac{1}{2}\right).
\end{align}
Under the $T$ transformation, the wavefunction with the WL in right-hand side of Eq.~(\ref{psiWL}) is transformed as
\begin{align}
\psi_{T^2}^{(j+0,0),M}\left(z+\widetilde{a}_w+\frac{1}{2},\tau+1\right) = e^{\pi ij} e^{\pi i\frac{j^2}{M}} e^{\frac{\pi i}{2}\frac{{\rm Im}(Mz+\alpha_1\tau-\alpha_2)}{{\rm Im}\tau}} \psi_{T^2}^{(j+0,0),M}(z+\widetilde{a}_w,\tau).
\label{psiWLT}
\end{align}
On the other hand, in this case, the gauge phase in Eq.~(\ref{SStoWL}) is also transformed:
\begin{align}
\widetilde{\psi}^{\alpha_1,\alpha_2}(z,\tau+1) &= e^{\frac{\pi i}{2}M\frac{{\rm Im}z}{{\rm Im}\tau}} e^{\pi i\frac{{\rm Im}(\alpha_1\bar{\tau}-\alpha_2)z}{{\rm Im}\tau}} \psi^{\alpha_1,\alpha_2}(z,\tau+1).
\label{SStoWLT}
\end{align}
Considering this equation and Eq.~(\ref{rhoTSS}), actually, the $T$ transformation for the wavefunction in left-hand side of Eq.~(\ref{psiWL}) is consistent with Eq.~(\ref{psiWLT}).

\section{$\mathbb{Z}_N$ Scherk-Schwarz phases and $\mathbb{Z}_N$ shift modes}
\label{SSshift}

Here, we also show that the wavefunctions on magnetized $T^2 \simeq \mathbb{C}/\Lambda$ with the $\mathbb{Z}_N$ SS phases are related to the $\mathbb{Z}_N$-eigenmode wavefunctions on magnetized full $\mathbb{Z}_N$ shifted orbifold of $\widetilde{T}^2 \simeq \mathbb{C}/\widetilde{\Lambda}\ (\widetilde{\Lambda}=N\Lambda)$ without the SS phases as follows.

First, the lattice vectors $\widetilde{e}_k\ (k=1,2)$ of the lattice $\widetilde{\Lambda}=N\Lambda$ are written by ones of the lattice $\Lambda$, $e_k\ (k=1,2)$, as $\widetilde{e}_k=Ne_k$.
Then, the coordinate and the modulus of $\widetilde{T}^2 \simeq \mathbb{C}/\widetilde{\Lambda}$, $(\widetilde{z},\widetilde{\tau}) \equiv (u/\widetilde{e}_1, \widetilde{e}_2/\widetilde{e}_1)$ are related to ones of $T^2 \simeq \mathbb{C}/\Lambda$, $(z,\tau) \equiv (u/e_1, e_2/e_1)$, as $(\widetilde{z},\widetilde{\tau}) = (z/N,\tau)$, where $u$ is the coordinate of $\mathbb{C}$.
Note that $\widetilde{z}+1 \sim \widetilde{z}$ and $\widetilde{z}+\widetilde{\tau} \sim \widetilde{z}$ are satisfied on $\widetilde{T}^2$.

The $\widetilde{T}^2/\mathbb{Z}_N$ full shifted orbifold~\cite{Kikuchi:2020frp}, on which the full modular symmetry remains, can be obtained by furthre identifying any $\mathbb{Z}_N$ shifted points $\widetilde{z}+(r+s\widetilde{\tau})/N\ (\forall r,s \in \mathbb{Z}_N)$ with $\widetilde{z}$. (See also Ref.~\cite{Fujimoto:2013xha}.)
Then, the boundary conditions of the wavefunction on the $\widetilde{T}^2/\mathbb{Z}_N$ full shifted orbifold with the magnetic flux $\widetilde{M}$ and the vanishing SS phases are just the following two conditions,
\begin{align}
\psi_{\widetilde{T}^2/\mathbb{Z}_N^{(\ell_1,\ell_2)}}\left(\widetilde{z}+\frac{1}{N},\widetilde{\tau}\right) &= e^{2\pi i\frac{\ell_1}{N}} e^{\pi i\widetilde{M}\frac{{\rm Im}\frac{\widetilde{z}}{N}}{{\rm Im}\widetilde{\tau}}} \psi_{\widetilde{T}^2/\mathbb{Z}_N^{(\ell_1,\ell_2)}}(\widetilde{z},\widetilde{\tau}), \label{z1N} \\
\psi_{\widetilde{T}^2/\mathbb{Z}_N^{(\ell_1,\ell_2)}}\left(\widetilde{z}+\frac{\widetilde{\tau}}{N},\widetilde{\tau}\right) &= e^{2\pi i\frac{\ell_2}{N}} e^{\pi i\widetilde{M}\frac{{\rm Im}\frac{\bar{\widetilde{\tau}}}{N}\widetilde{z}}{{\rm Im}\widetilde{\tau}}} \psi_{\widetilde{T}^2/\mathbb{Z}_N^{(\ell_1,\ell_2)}}(\widetilde{z},\widetilde{\tau}), \label{ztauN}
\end{align}
where $\ell_1, \ell_2 \in \mathbb{Z}_N$ are the $\mathbb{Z}_N$-eigenvalues.
From the above boundary conditions, $\widetilde{M}/N^2 \equiv M \in \mathbb{Z}$ should be satisfied.
The above wavefunction on the magnetized $\widetilde{T}^2/\mathbb{Z}_N$ full shifted orbifold without the SS phases, $\psi_{\widetilde{T}^2/\mathbb{Z}_N^{(\ell_1,\ell_2)}}^{j,M}$, can be expanded by the wavefunction on the magnetized $\widetilde{T}^2$ without the SS phases as
\begin{align}
\psi_{\widetilde{T}^2/\mathbb{Z}_N^{(\ell_1,\ell_2)}}^{j,M}(\widetilde{z},\widetilde{\tau})=\frac{1}{\sqrt{N}}\sum_{k=0}^{N-1}e^{-2\pi ik\frac{\ell_2}{N}}\psi_{\widetilde{T}^2}^{(Nj+\ell_1)+kNM,N^2M}(\widetilde{z},\widetilde{\tau}). \label{shift}
\end{align}

Furthermore, by considering the relation, $(\widetilde{z},\widetilde{\tau}) = (z/N,\tau)$, the boundary conditions in Eqs.~(\ref{z1N}) and (\ref{ztauN}) correspond to ones in Eqs.~(\ref{psiz1SS}) and (\ref{psiztauSS}) with the $\mathbb{Z}_N$ SS phases, $(\alpha_1,\alpha_2)=(\ell_1/N, \ell_2/N)\ (\ell_1,\ell_2 \in \mathbb{Z}_N)$.
Actually, the above wavefunction with the $\mathbb{Z}_N$-eigenvalue, $(\ell_1,\ell_2)$, on the $\widetilde{T}^2/\mathbb{Z}_N$ full shifted orbifold with the magnetic flux $\widetilde{M}$ and the vanishing SS phases is related to the wavefunction on $T^2$ with the magnetic flux $M$ and the $\mathbb{Z}_N$ SS phases, $(\alpha_1,\alpha_2)=(\ell_1/N, \ell_2/N)$, as
\begin{align}
\psi_{\widetilde{T}^2/\mathbb{Z}_N^{(\ell_1,\ell_2)}}^{j,M}\left(\frac{z}{N},\tau\right)
&=\frac{1}{\sqrt{N}}\sum_{k=0}^{N-1}e^{-2\pi ik\frac{\ell_2}{N}}\psi_{\widetilde{T}^2}^{(Nj+\ell_1)+kNM,N^2M}\left(\frac{z}{N},\tau\right) \notag \\
&=e^{2\pi i \left(j+\frac{\ell_1}{N}\right)\frac{\ell_2}{N}/M}\psi_{T^2}^{\left(j+\frac{\ell_1}{N},\frac{\ell_2}{N}\right),M}(z,\tau). \label{shiftSS}
\end{align}
The analyses of the modular transformation are also consistent.

Similarly, the wavefunction on the magnetized $\widetilde{T}^2/\mathbb{Z}_2$ twisted and full shifted orbifold without the SS phases is related to one on the magnetized $T^2$ with the $\mathbb{Z}_2$ SS phases.
Their behavior of the modular transformation are consistent each other.


\section{$\widetilde{\Delta}(6M^2)$ as subgroup of $\widetilde{\Gamma}_{2M}$}
\label{proof}

Here, we prove the generators in Eq.~(\ref{DeltatildeST}), in particular for $M \in 4\mathbb{Z}$, satisfy the algebraic relations of $\widetilde{\Delta}(6M^2)$ in Eq.~(\ref{algDeltatilde}), where the algebraic relations of $\widetilde{\Gamma}_{2M}$ in Eqs.~(\ref{Z})-(\ref{TN}) with $N=2M$ and also the additional relation in Eq.~(\ref{X}) are satisfied.
Note that when we have $k/2=$integer [even] in Eqs.~(\ref{Z})-(\ref{TN}) with $k=$integer [even] and $N=2M$, which correspond to the algebraic relations of $\Gamma'_{2M}$ [$\Gamma_{2M}$] in Eqs.~(\ref{Gammarep}) and (\ref{Gamma'Nrep}) with $N=2M$, we can find that the generators in Eq.~(\ref{DeltatildeST}) corresponds to ones in Eq.~(\ref{Delta'ST}) [Eq.~(\ref{DeltaST})] and they satisfy the algebraic relations of $\Delta'(6M^2)$ [$\Delta(6M^2)$] in Eq.~(\ref{algDelta'}) [Eq.~(\ref{algDelta})].

First, by using Eqs.~(\ref{Z})-(\ref{ZT}), Eq.~(\ref{X}) can be rewritten\footnote{When we consider Eqs.~(\ref{Gammarep}) and (\ref{Gamma'Nrep}) with $N=2M, \ M=1,2$, we can check that Eq.~(\ref{reX}) is already satisfied.} as
\begin{align}
(S^7T^3)^3 = (S^{-1}T^3)^3 = \mathbf{1}. \label{reX}
\end{align}
By using Eqs.~(\ref{Z})-(\ref{ZT}), and (\ref{reX}), the generator $a'$ in Eq.~(\ref{DeltatildeST}) can be rewritten as
\begin{align}
a'
&= ST^2S^{-1}T^{-2} \notag \\
&= ST TS^{-1}T^2 T^{-4} \notag \\
&= ST T^{-2}ST^{-1} T^{-2}ST^{-1} T^{-4} \notag \\
&= S^{-1}T^{-1}S^{-1}T^{-1} S^4 T^{-2}ST^{-5} \notag \\
&= TS T^{-2}ST^{-5} \notag \\
&= TS^{-1}T^{-1}S^4T^{-1}S^{-1}T^{-5} \notag \\
&= T^2STS S^4 STST^{-4} \notag \\
&= T^2ST^2S^{-1}T^{-4} \notag \\
&= T^2 (ST^2S^{-1}T^{-2}) T^{-2} \notag \\
\Leftrightarrow \ 
a'
&= T^{-2} (ST^2ST^{-2}) T^2 \notag \\
&= T^{-2}ST^2S^{-1}. \label{a're}
\end{align}
Then, we can obtain
\begin{align}
ST^{2p}S^{-1}T^{2q} =(ST^2S^{-1})^{p}T^{2q} = T^{2q}ST^{2p}S^{-1}, \quad p,q \in \mathbb{Z},  \label{T2p2q}
\end{align}
in general.
Similarly, by using this relation, the generator $a$ in Eq.~(\ref{DeltatildeST}) can be rewritten as
\begin{align}
a
&= ST^2S^5T^4 \notag \\
&= T^4ST^2S^5. \label{are}
\end{align}
Thus, we can obtain
\begin{align}
&a^M = S^{-2M}T^{4M}ST^{2M}S^{-1} = 1, \quad a'^M = T^{-2M}ST^{2M}S^{-1} = 1, \label{aa'M} \\
&aa' = ST^4S^5T^2 = a'a, \label{aa'a'a}
\end{align}
by also using Eq.~(\ref{TN}) with $N=2M$ and $M \in 4\mathbb{Z}$~\footnote{It is because $S^{-2M}=1$ is satisfied only if $M \in 4\mathbb{Z}$. However, when we consider the case that Eqs.~(\ref{Gammarep}) and (\ref{Gamma'Nrep}) with $N=2M$ are satisfied instead of Eqs.~(\ref{Z})-(\ref{TN}) with $N=2M$, $S^{-2M}=1$ is satisfied even if $M=2(2s-1)\ (s \in \mathbb{Z})$.}.
Furthermore, from Eq.~(\ref{Z2}), we also have
\begin{align}
(S^5T)^3=\mathbf{1}.
\end{align}
Then, we can prove
\begin{align}
(S^{2n+3}T^{2n-1})^3=\mathbf{1}, \ n \in \mathbb{N},
\end{align}
on the mathematical induction.
Thus, we can obtain the other relations in Eq.~(\ref{algDeltatilde}),
\begin{align}
b^3 &= T^{-M} (T^{\frac{3}{2}M+3}S^{\frac{3}{2}M-1})^3 T^M = 1 \label{b3} \\
c^2
&= ST^{M-2}ST^{\frac{3}{2}M-1} ST^{M-2}ST^{\frac{3}{2}M-1} \notag \\
&= ST^{M-2}S^{-1}T^{-1}S^{-1}T^{M-2}ST^{M-1}S^4 \notag \\
&= ST^{M-1}ST^{M-1}ST^{M-1} \notag \\
&= (S^{M+3}T^{M-1})^3 S^{-3M-6} \notag \\
&= S^{M+2} \label{c2} \\
c^4 &= S^4 \label{c4} \\
c^8 &= 1 \label{c8} \\
cbc^{-1}
&= ST^{M-2}ST^2S^{\frac{3}{2}M-1}T^{1-\frac{M}{2}}S^{-1}T^{2-M}S^{-1} \notag \\
&= T^2 ST^{M-2}S^{\frac{3}{2}M} T^5S^{-1}T^{2-M}S^{-1}T^{-\frac{M}{2}-4} \notag \\
&= T^2 ST^{M+3}S^{\frac{3}{2}M-1} T^{3-M}S^5TS T^{-\frac{M}{2}-3} \notag \\
&= T^2 ST^{M+3}S^{M-1}T^{M+3}S^5TS^{-\frac{3}{2}M+1} T^{-\frac{M}{2}-3} \notag \\
&= T^2 S^{M-1}T^{M+3}S^{M-1}T^{M+3} S^{-M-1}T S^{-\frac{3}{2}M+1} T^{-\frac{M}{2}-3} \notag \\
&= T^2 T^{-M-3} S^{-M+1} S^{-M-1} T S^{-\frac{3}{2}M+1} T^{-\frac{M}{2}-3} \notag \\
&= T^{-M} S^{-\frac{3}{2}M+1} T^{-\frac{M}{2}-3} \notag \\
&= b^{-1} \label{cbc-1b-1}
\end{align}
\begin{align}
bab^{-1}
&= T^{\frac{M}{2}+3} S^{\frac{3}{2}M-1} T^{M} ST^2S^5T^{4-M} S^{-\frac{3}{2}M+1} T^{-\frac{M}{2}-3} \notag \\
&= T^{\frac{M}{2}+3} S^{\frac{3}{2}M-1} T^{4} S^{-\frac{3}{2}M-1} T^{-\frac{M}{2}-1} \notag \\
&= T S^{-1}T^4S^{-1}T \notag \\
&= T^{-2}S^{-1}T^{-3} STST \notag \\
&= T^{-2}S^{-5}T^{-4}S^{-1} \notag \\
&= a^{-1}a'^{-1} \label{bab-1a-1a'-1} \\
ba'b^{-1}
&= T^{\frac{M}{2}+3} S^{\frac{3}{2}M-1} T^{M} ST^2S^{-1}T^{-2-M} S^{-\frac{3}{2}M+1} T^{-\frac{M}{2}-3} \notag \\
&= T^{\frac{M}{2}+3} S^{\frac{3}{2}M-1} T^{-2} S^{-\frac{3}{2}M+1} T^{-\frac{M}{2}-1} \notag \\
&= T^{-1} S^{-1} T^{-2} S^{-1} T^{-1} S^{2} T^{4} \notag \\
&= STS S^2 STS T^4 \notag \\
&= ST^2S^5T^4 \notag \\
&= a \label{ba'-1ba} \\
cac^{-1}
&= ST^{M-2}ST^{\frac{3}{2}M-1} ST^2S^5T^{5-\frac{3}{2}M} S^{-1} T^{2-M} S^{-1} \notag \\
&= ST^{M-2}S^{-1}T^{-1}S^{-1}T^2S^5T^5ST^{2-M}S \notag \\
&= ST^{M-1}S^5T^3S^5T^5ST^{2-M}S \notag \\
&= S^{-1}T^{M-2} TS^7T^3S^7T^2 T^3 ST^{2-M}S^{-1} \notag \\
&= S^{-1}T^{M-4} S T^2 ST^{2-M}S^{-1} \notag \\
&= T^2ST^{-2}S^{-1} \notag \\
&= a'^{-1} \label{cac-1a'-1} \\
ca'c^{-1}
&= ST^{M-2}ST^{\frac{3}{2}M-1} ST^2S^{-1}T^{-1-\frac{3}{2}M} S^{-1} T^{2-M} S^{-1} \notag \\
&= ST^{M-2}S^{-1}T^{-1} ST^2S^{-1}T^{-1} S^{-1} T^{2-M} S \notag \\
&= ST^{M-1}S^{-1}T^4S^5T^{3-M}S \notag \\
&= STST^4S^3TS \notag \\
&= T^{-1}S^3T^3S^3TS \notag \\
&= T^{-1}S^5T^2S^{3}T^{-1} \notag \\
&= T^{-2}S^{-1}T^{-1}S^{-1}S^6S^{-1}T^{-1}S^{-1}T^{-2} \notag \\
&= T^{-2} S^{-5} T^{-2} S^{-1} T^{-2} \notag \\
&= T^{-4} S^{-5} T^{-2} S^{-1} \notag \\
&= a^{-1} \label{ca'c-1a-1}.
\end{align}
Therefore, when the relation in Eq.~(\ref{X}) is also satisfied in addition to the algebraic relations of $\widetilde{\Gamma}_{2M}$, in particular for $M \in 4\mathbb{Z}$, Eq.~(\ref{DeltatildeST}) can be the generators of $\widetilde{\Delta}(6M^2)$.
Similarly, when the algebraic relations of $\Gamma'_{2M}$ [$\Gamma_{2M}$] and also Eq.~(\ref{X}) are satisfied, we can find that Eq.~(\ref{Delta'ST}) [Eq.~(\ref{DeltaST})] as well as Eq.~(\ref{Delta'STodd}) [the generators in footnote 8] can be the generators of $\Delta'(6M^2)$ [$\Delta(6M^2)$].


\section{Three-dimensional modular forms}
\label{3dmodularform}

Here, we express three-dimensional modular forms obtained from the wavefunctions on magnetized orbifolds at $z=0$, which means the modular forms can be obtained from $\mathbb{Z}_2$-even ($m=n=0$) modes.

We can obtain two three-dimensional modular forms of weight $1/2$ obtained from the modes $(M,m)=(4,0)$ and $(7,0)$ at $z=0$ on the magnetized $T^2/\mathbb{Z}_2^{{\rm(t)}}$ twisted orbifold, as the followings,
\begin{align}
\begin{pmatrix}
\vartheta^{4}_{1}(\tau) \\ \vartheta^{4}_{2}(\tau) \\ \vartheta^{4}_{3}(\tau)
\end{pmatrix}
&=
\begin{pmatrix}
\vartheta
\begin{bmatrix}
0 \\ 0
\end{bmatrix}
(0, 4\tau) \\
\frac{1}{\sqrt{2}} \left(
\vartheta
\begin{bmatrix}
\frac{1}{4} \\ 0
\end{bmatrix}
(0, 4\tau)
+
\vartheta
\begin{bmatrix}
\frac{3}{4} \\ 0
\end{bmatrix}
(0, 4\tau) \right) \\
\vartheta
\begin{bmatrix}
\frac{2}{4} \\ 0
\end{bmatrix}
(0, 4\tau)
\end{pmatrix}
=
\begin{pmatrix}
\vartheta
\begin{bmatrix}
0 \\ 0
\end{bmatrix}
(0, 4\tau) \\
\sqrt{2}
\vartheta
\begin{bmatrix}
\frac{1}{4} \\ 0
\end{bmatrix}
(0, 4\tau) \\
\vartheta
\begin{bmatrix}
\frac{2}{4} \\ 0
\end{bmatrix}
(0, 4\tau)
\end{pmatrix}
\quad  (M=4) \\
\begin{pmatrix}
\vartheta^{7}_{1}(\tau) \\ \vartheta^{7}_{2}(\tau) \\ \vartheta^{7}_{3}(\tau)
\end{pmatrix}
&=
\begin{pmatrix}
\frac{1}{\sqrt{2}} \left(
\vartheta
\begin{bmatrix}
\frac{1}{14} \\ -\frac{1}{2}
\end{bmatrix}
(0, 7\tau)
-
\vartheta
\begin{bmatrix}
\frac{13}{14} \\ -\frac{1}{2}
\end{bmatrix}
(0, 7\tau) \right) \\
\frac{1}{\sqrt{2}} \left(
\vartheta
\begin{bmatrix}
\frac{3}{14} \\ -\frac{1}{2}
\end{bmatrix}
(0, 7\tau)
-
\vartheta
\begin{bmatrix}
\frac{11}{14} \\ -\frac{1}{2}
\end{bmatrix}
(0, 7\tau) \right) \\
\frac{1}{\sqrt{2}} \left(
\vartheta
\begin{bmatrix}
\frac{5}{14} \\ -\frac{1}{2}
\end{bmatrix}
(0, 7\tau)
-
\vartheta
\begin{bmatrix}
\frac{9}{14} \\ -\frac{1}{2}
\end{bmatrix}
(0, 7\tau) \right)
\end{pmatrix}
= 
\begin{pmatrix}
\sqrt{2}
\vartheta
\begin{bmatrix}
\frac{1}{14} \\ -\frac{1}{2}
\end{bmatrix}
(0, 7\tau) \\
\sqrt{2}
\vartheta
\begin{bmatrix}
\frac{3}{14} \\ -\frac{1}{2}
\end{bmatrix}
(0, 7\tau) \\
\sqrt{2}
\vartheta
\begin{bmatrix}
\frac{5}{14} \\ -\frac{1}{2}
\end{bmatrix}
(0, 7\tau)
\end{pmatrix}
\ (M=7).
\end{align}
They are the modular forms of weight $1/2$ for $\widetilde{\Gamma}(8)$ and $\widetilde{\Gamma}(56)$, respectively, and also they transform as the three-dimensional representations of $\widetilde{\Delta}(96)$ and $PSL(2,Z_7) \times Z_8$, respectively.

Similarly, we can obtain four three-dimensional modular forms of weight $1$, two of which are obtained from the modes $(M^{(1)},m_1:M^{(2)},m_2)=(4,0:3,0)$ and $(7,0:3,0)$ at $z_1=z_2=0$ on the magnetized $T^2_1/\mathbb{Z}_2^{({\rm t}_1)} \times T^2/\mathbb{Z}_2^{({\rm t}_2)}$ orbifold, and the other two of which are obtained from the modes $(M,m,n)=(2,0,0)$ and $(5,0,0)$ at $z_1=z_2=0$ on the magnetized $(T^2_1 \times T^2_2)/(\mathbb{Z}_2^{\rm (t)}\times \mathbb{Z}_2^{\rm (p)})$ orbifold as the followings,
\begin{align}
\begin{pmatrix}
\vartheta^{(4,3)}_{1}(\tau) \\ \vartheta^{(4,3)}_{2}(\tau) \\ \vartheta^{(4,3)}_{3}(\tau)
\end{pmatrix}
&=
\begin{pmatrix}
\sqrt{2}
\vartheta
\begin{bmatrix}
\frac{1}{6} \\ -\frac{1}{2}
\end{bmatrix}
(0, 3\tau)
\vartheta
\begin{bmatrix}
0 \\ 0
\end{bmatrix}
(0, 4\tau) \\
2
\vartheta
\begin{bmatrix}
\frac{1}{6} \\ -\frac{1}{2}
\end{bmatrix}
(0, 3\tau)
\vartheta
\begin{bmatrix}
\frac{1}{4} \\ 0
\end{bmatrix}
(0, 4\tau) \\
\sqrt{2}
\vartheta
\begin{bmatrix}
\frac{1}{6} \\ -\frac{1}{2}
\end{bmatrix}
(0, 3\tau)
\vartheta
\begin{bmatrix}
\frac{2}{4} \\ 0
\end{bmatrix}
(0, 4\tau)
\end{pmatrix} \quad (M^{(1)}=4,M^{(2)}=3) \\
\begin{pmatrix}
\vartheta^{(7,3)}_{1}(\tau) \\ \vartheta^{(7,3)}_{2}(\tau) \\ \vartheta^{(7,3)}_{3}(\tau)
\end{pmatrix}
&=
\begin{pmatrix}
2
\vartheta
\begin{bmatrix}
\frac{1}{6} \\ -\frac{1}{2}
\end{bmatrix}
(0, 3\tau)
\vartheta
\begin{bmatrix}
\frac{1}{14} \\ -\frac{1}{2}
\end{bmatrix}
(0, 7\tau) \\
2
\vartheta
\begin{bmatrix}
\frac{1}{6} \\ -\frac{1}{2}
\end{bmatrix}
(0, 3\tau)
\vartheta
\begin{bmatrix}
\frac{3}{14} \\ -\frac{1}{2}
\end{bmatrix}
(0, 7\tau) \\
2
\vartheta
\begin{bmatrix}
\frac{1}{6} \\ -\frac{1}{2}
\end{bmatrix}
(0, 3\tau)
\vartheta
\begin{bmatrix}
\frac{5}{14} \\ -\frac{1}{2}
\end{bmatrix}
(0, 7\tau)
\end{pmatrix} \quad (M^{(1)}=7,M^{(2)}=3), \\
\begin{pmatrix}
\vartheta^{(2,2)}_{1}(\tau) \\ \vartheta^{(2,2)}_{2}(\tau) \\ \vartheta^{(2,2)}_{3}(\tau)
\end{pmatrix}
&=
\begin{pmatrix}
\vartheta
\begin{bmatrix}
0 \\ 0
\end{bmatrix}
(0, 2\tau)
\vartheta
\begin{bmatrix}
0 \\ 0
\end{bmatrix}
(0, 2\tau) \\
\sqrt{2}
\vartheta
\begin{bmatrix}
0 \\ 0
\end{bmatrix}
(0, 2\tau)
\vartheta
\begin{bmatrix}
\frac{1}{2} \\ 0
\end{bmatrix}
(0, 2\tau) \\
\vartheta
\begin{bmatrix}
\frac{1}{2} \\ 0
\end{bmatrix}
(0, 2\tau)
\vartheta
\begin{bmatrix}
\frac{1}{2} \\ 0
\end{bmatrix}
(0, 2\tau)
\end{pmatrix} \quad (M^{(1)}=M^{(2)}=M=2) \\
\begin{pmatrix}
\vartheta^{(5,5)}_{1}(\tau) \\ \vartheta^{(5,5)}_{2}(\tau) \\ \vartheta^{(5,5)}_{3}(\tau)
\end{pmatrix}
&=
\begin{pmatrix}
2
\vartheta
\begin{bmatrix}
\frac{1}{10} \\ -\frac{1}{2}
\end{bmatrix}
(0, 5\tau)
\vartheta
\begin{bmatrix}
\frac{1}{10} \\ -\frac{1}{2}
\end{bmatrix}
(0, 5\tau) \\
2\sqrt{2}
\vartheta
\begin{bmatrix}
\frac{1}{10} \\ -\frac{1}{2}
\end{bmatrix}
(0, 5\tau)
\vartheta
\begin{bmatrix}
\frac{3}{10} \\ -\frac{1}{2}
\end{bmatrix}
(0, 5\tau) \\
2
\vartheta
\begin{bmatrix}
\frac{3}{10} \\ -\frac{1}{2}
\end{bmatrix}
(0, 5\tau)
\vartheta
\begin{bmatrix}
\frac{3}{10} \\ -\frac{1}{2}
\end{bmatrix}
(0, 5\tau)
\end{pmatrix} \quad (M^{(1)}=M^{(2)}=M=5).
\end{align}
They are the modular forms of weight $1$ for $\Gamma(24)$, $\Gamma(84)$, $\Gamma(4)$, and $\Gamma(20)$ respectively, and also they transform as the three-dimensional representations of $\Delta'(96) \times Z_3$, $PSL(2,Z_7) \times Z_3 \times Z_4$, $S'_4$, and $A_5 \times Z_4$, respectively.


\end{document}